\documentclass[aoas]{imsart}
\usepackage{amsmath,natbib,amssymb, amsthm}
\usepackage{bm} \usepackage{bbm}
\usepackage{float}
\usepackage[normalem]{ulem}
\usepackage{url}
\usepackage{multirow}

\newcommand{\bbR}{\mathbb{R}} \newcommand{\cC}{\mathcal{C}}

 \newcommand{\0}{{\mathbf{0}}}

\newcommand{\bu}{{\mathbf{u}}} 
\newcommand{\x}{{\mathbf{x}}}

 \newcommand{\E}{{\mathbf{E}}}

 \renewcommand{\P}{{\mathbf{P}}}

\newcommand{\1}{{\mathbf{1}}}
\newcommand{\indep}{\perp\!\!\!\perp}

\newcommand{\bx}{{\mathbf{x}}} 
\newcommand{\bR}{{\mathbf{R}}} 
\newcommand{\bZ}{{\mathbf{Z}}}
\newcommand{\bz}{{\mathbf{z}}}  
\newcommand{\cF}{{\mathcal{F}}} 
 
\newcommand{\bF}{{\mathbf{F}}} 
\newcommand{\bof}{{\mathbf{f}}} 
\newcommand{\bq}{{\mathbf{q}}}
\newcommand{\bpi}{{\boldsymbol\pi}}
\newcommand{\var}{\text{var}}
\renewcommand{\E}{{\mathbb{E}}}
\renewcommand{\P}{{\mathbb{P}}}

\newcommand{\cN}{{\cal N}}
\newcommand{\cI}{{\cal I}}

\usepackage{verbatim} 
\usepackage{enumitem} 
\usepackage{appendix} \usepackage{pdfpages}
\usepackage[usenames,dvipsnames]{pstricks} \usepackage{epsfig}
\usepackage{pst-grad} 

\newcommand*{\Chi}{\mbox{\large$\chi$}} 

\newtheorem{theorem}{Theorem}

\newtheorem{proposition}[theorem]{Proposition}

\usepackage{fancyhdr}
\begin{document}

\begin{frontmatter}

  \title{Extended Sensitivity Analysis for Heterogeneous Unmeasured Confounding with An Application to Sibling Studies of Returns to Education}
  \runtitle{Extended Sensitivity Analysis for Heterogeneous Bias}

  \begin{aug}
    \author{\fnms{Colin B.} \snm{Fogarty}\thanksref{a1,eq1}\ead[label=e1]{cfogarty@mit.edu}}
    \and
    \author{\fnms{Raiden B.} \snm{Hasegawa}\thanksref{a2}\ead[label=e2]{raiden@wharton.upenn.edu}}
    \thankstext{eq1}{Both authors contributed equally.}
    \runauthor{C. B. Fogarty and R. B. Hasegawa}
    
    \affiliation{Massachusetts Institute of Technology\thanksmark{a1} and University of Pennsylvania\thanksmark{a2}}

    \address{C. B. Fogarty \\
     Operations Research and Statistics Group \\
      Sloan School of Management \\
      Massachusetts Institute of Technology\\
      Cambridge, Massachusetts 02142 \\
      USA \\
      \printead{e1}
    }

    \address{R. B. Hasegawa \\
      Department of Statistics \\
      The Wharton School \\
      University of Pennsylvania \\
      Philadelphia, Pennsylvania 19104 \\
      USA \\
      \printead{e2}
    }
  \end{aug}

  \begin{abstract} The conventional model for assessing insensitivity to hidden bias in paired observational studies constructs a worst-case distribution for treatment assignments subject to bounds on the maximal bias to which any given pair is subjected. In studies where rare cases of extreme hidden bias are suspected, the maximal bias may be substantially larger than the typical bias across pairs, such that a correctly specified bound on the maximal bias would yield an unduly pessimistic perception of the study's robustness to hidden bias. We present an extended sensitivity analysis which allows researchers to simultaneously bound the maximal and typical bias perturbing the pairs under investigation while maintaining the desired Type I error rate. We motivate and illustrate our method with two sibling studies on the impact of schooling on earnings, one containing information of cognitive ability of siblings and the other not. Cognitive ability, clearly influential of both earnings and degree of schooling, is likely similar between members of most sibling pairs yet could, conceivably, vary drastically for some siblings. The method is straightforward to implement, simply requiring the solution to a quadratic program. \texttt{R} code is provided at the author's website \url{http://www.raidenhasegawa.com} 
\end{abstract}


\end{frontmatter}

\section{Introduction}
\subsection{A motivating example: Returns to schooling}
Is educational attainment a determining factor for success in the labor market? Initial interest among economists in addressing this question is attributed to the observation in the late 1950s that increases in education levels could account for much of the productivity growth in post-war US \citep{bec64, gri70, car99}. With strong evidence of a positive association between education and earnings in a variety of political and geographic environments but little to no experimental data, a recurring theme in the subsequent pursuit of a causal relationship between education and income is that of the presence of ``ability bias'' \citep{car99}. After controlling for family background, or considering within-family estimates of the causal effect using sibling or twin studies, can latent differences in ability influence both differences in schooling choice and earnings? A notable twin study by \citet{ashenfelter1998}, which we re-examine in this paper, argued cogently, albeit with limited statistical evidence, that identical twins can be regarded as truly identical in all dimensions relevant to schooling choices and future income, including latent ability. In a survey of contemporary economic investigations of returns to education, \citet[][p.1852]{car99} addresses this hypothesis:

\begin{quote}
  Despite this evidence, and the strong intuitive appeal of the ``equal abilities'' assumption for identical twins, however, I suspect that observers with a strong a priori belief in the importance of ability bias will remain unconvinced. 
\end{quote}

Perhaps latent ability is truly identical for many twin pairs but markedly different in a few pairs; what would happen then? That exogeneity is not testable leaves even the most compelling observational evidence susceptible to the warranted, though often non-specific, criticism, ``what if bias remains?" Should the totality of evidence assume the absence of hidden bias, the critic need merely suggest the existence of bias to cast doubt upon the posited causal mechanism. It is thus incumbent upon researchers not only to anticipate such criticism, but also to arm themselves with a suitable rejoinder. Rather than arguing for or against the presence of ability bias or any other unobserved confounding factor, in this paper we assess the sensitivity of causal conclusions to departures from truly randomized assignment while allowing for patterns of ability bias that may be highly heterogeneous across sibling pairs.


\subsection{Assessing returns to schooling with sibling comparison designs}\label{sec.sibComp}
Sibling comparison studies are a special case of stratified designs where natural blocks are formed by family membership. These studies automatically control for genetic, socioeconomic, cultural, and child-rearing characteristics to the extent that they are shared between siblings; however, instability of familial characteristics over time for sibling pairs of different ages and non-shared genetic makeup are among threats to this premise \citep{donovan2011}. Due to their natural and automatic control of stable familial factors, both observed and unobserved, sibling comparison designs have long been a popular tool for studying causal effects in both epidemiological and economic settings; see \citet{griliches1979} and \citet{donovan2011} for surveys of past and current sibling comparison studies in economics and epidemiology, respectively.

Sibling comparison designs have been particularly fruitful in the study of returns to schooling, where genetic and family background are deemed essential to both schooling choices and future income; see for example \citet{hauser1999}, \citet{stanek2011}, and \citet{ashenfelter1998}. \citet{hauser1999} study sibling pairs from the Wisconsin Longitudinal Study (WLS), a random sample ($n=10,317$) of men and women born between 1938 and 1940 who graduated from Wisconsin high schools in 1957. The size of the sample was set to be approximately a third of all Wisconsin high school graduates in 1957. Random siblings of those in the study ($n = 7,928$), born between 1930 and 1948, were also selected and interviewed. The WLS contains a rich set of baseline covariates and endpoints, including physical, cognitive, social, and occupational outcomes collected over nearly 60 years following graduation. Uniquely, the WLS dataset contains intelligence quotient (IQ) scores recorded while a given individual was in high school -- a covariate rarely measured in longitudinal cohort studies. 

\begin{figure}[h]
  \centering
  \begin{tabular}{c}
    \includegraphics[scale=0.5]{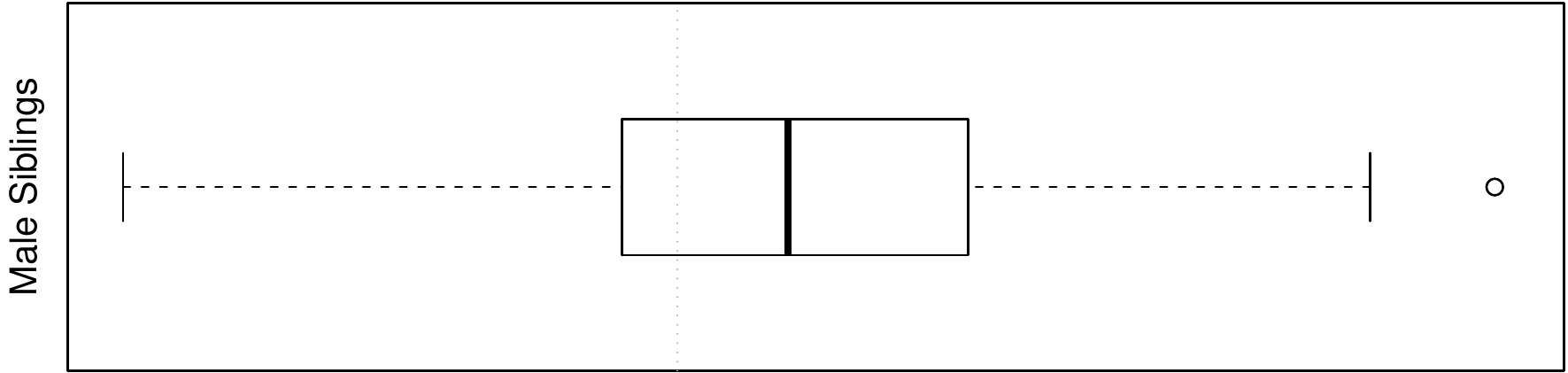}\vspace{.2cm}\\
    \includegraphics[scale=0.5]{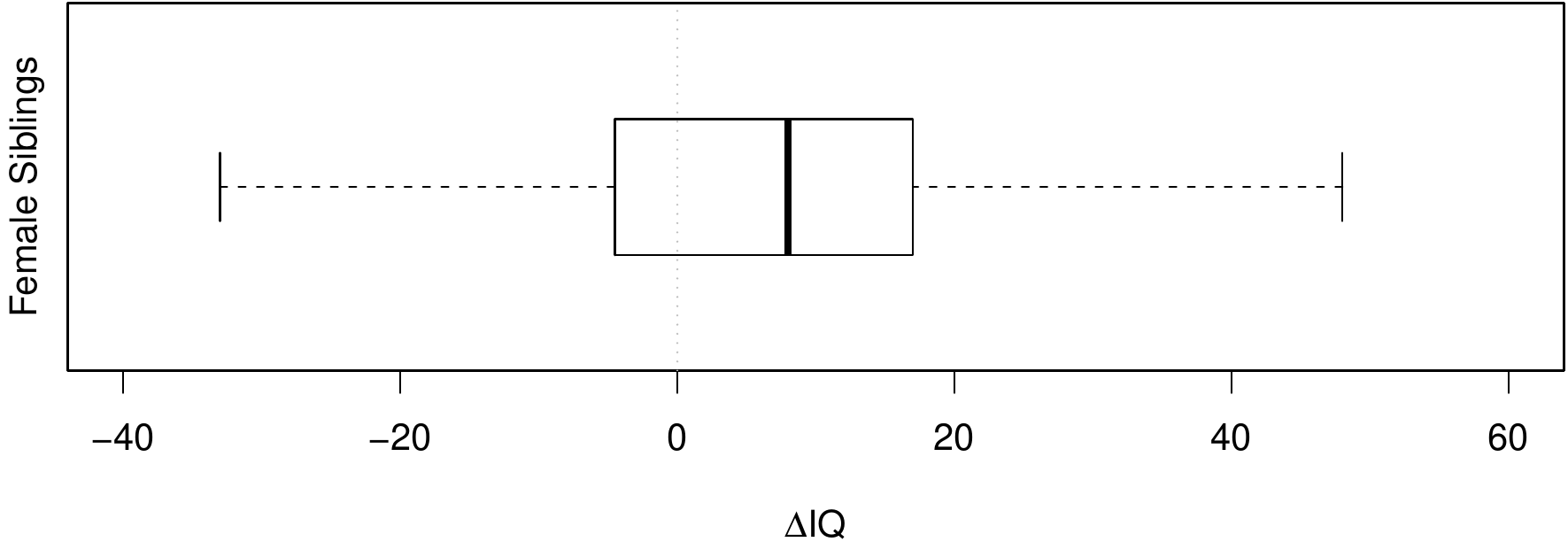}
  \end{tabular} 
  \caption{Boxplots of differences in IQ scores between same-sex siblings where one attended college and the other did not. \textit{(top panel):} Male same-sex sibling pairs ($n=128$). \textit{(bottom panel):} Female same-sex sibling pairs ($n=43$).}  \label{fig.boxiq}
\end{figure}

In other sibling studies of the returns to schooling, such as that of \citet{ashenfelter1998}, baseline intelligence measures such as IQ are not available, making it plausible that the siblings being compared differ in cognitive ability in unobserved ways. Furthermore, the IQ data from the WLS study suggests that, when considering same-sex sibling pairs where one sibling attended college and the other did not ($n=171$), intellectual ability is not balanced sufficiently by shared genetics alone. The boxplots of differences in IQ between the college-attending siblings and their counterparts in Figure \ref{fig.boxiq} exhibit a prominent shift in the IQ distribution between the two groups for both male and female same-sex sibling pairs. The mean (sd) is $107.1\,(14.7)$ in the college-attending group and $97.4\,(14.4)$ in the high school-only group for male same-sex sibling pairs. In female same-sex sibling pairs, these values are $108.1\,(14.0)$ and $101.4\,(14.2)$ for the college-attending and high school-only attending groups respectively. Details on the construction of the $171$ same-sex sibling pairs can be found in Appendix \ref{sec:A2}. An important inclusion criterion was that both siblings were employed when income data was collected.

\subsection{Potential for rare but extreme unmeasured biases}
Despite their analytical strengths and convenient, automatic stratification, sibling comparison designs for estimating causal effects are subject to biases arising from differences in subject-level confounders. For example, latent ability, as measured by IQ, may differ substantially within twin pairs in Ashenfelter and Rouse's twin study. This concern is magnified in sibling studies where discordant within-pair treatment assignment may actually exacerbate differences in covariates that are related to both the intervention and outcome of interest \citep{frisell2012}. When pairs do not arise naturally, as in paired sibling studies, matching algorithms designed to minimize disparities in observed covariates may be used to construct pairs of ``comparable'' subjects; see, for example, \citet{han06} and \citet{stu10} for discussion on various approaches to matching. Matched pairs constructed in this fashion may be comparable along observed covariates, but they are still vulnerable to unmeasured bias arising from differences in covariates not available to the matching algorithm.  

While agnostic covariate adjustment within sibling sets as suggested in \citet{ros02cov} can help mitigate the impact of discrepancies in observed individual-specific covariates, bias arising from differences in unobserved confounders may remain and imperil the conclusions of the study. An additional inferential step known as a \textit{sensitivity analysis} assesses the robustness of the conclusions of a study to these unmeasured biases. Sensitivity analysis was first introduced by \citet{cor59} and refined to accommodate continuous outcomes in \citet{rosenbaum1987}. The resulting sensitivity analysis for paired studies considers the worst-case bias to which any pair may be subject and asks whether the study conclusions might change if we assumed that \textit{all} pairs were exposed to the maximal bias in a manner adverse to the desired inference. We refer to this as the \textit{conventional} sensitivity analysis. See \citet{cor59}, \citet{mar97}, \citet{imb03}, \citet{yu05}, \citet{wang2006}, \citet{egl09}, \citet{hos10}, \citet{zub13}, \citet{liu13}, and \citet{van17} for additional perspectives on and worked examples of sensitivity analysis. 

\begin{figure}
  \centering
  \begin{tabular}{cc}
    \includegraphics[scale=0.3]{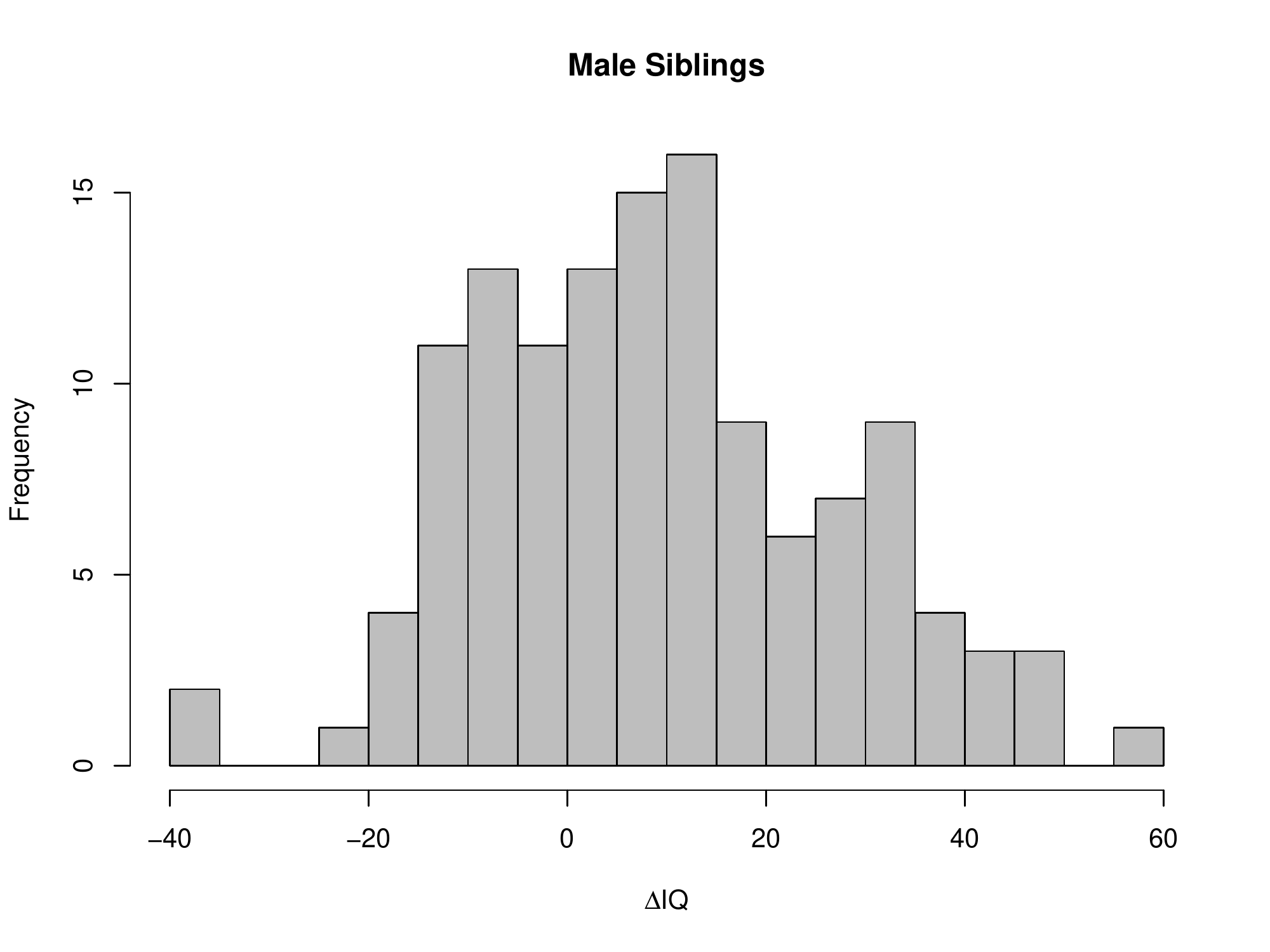} & \includegraphics[scale=0.3]{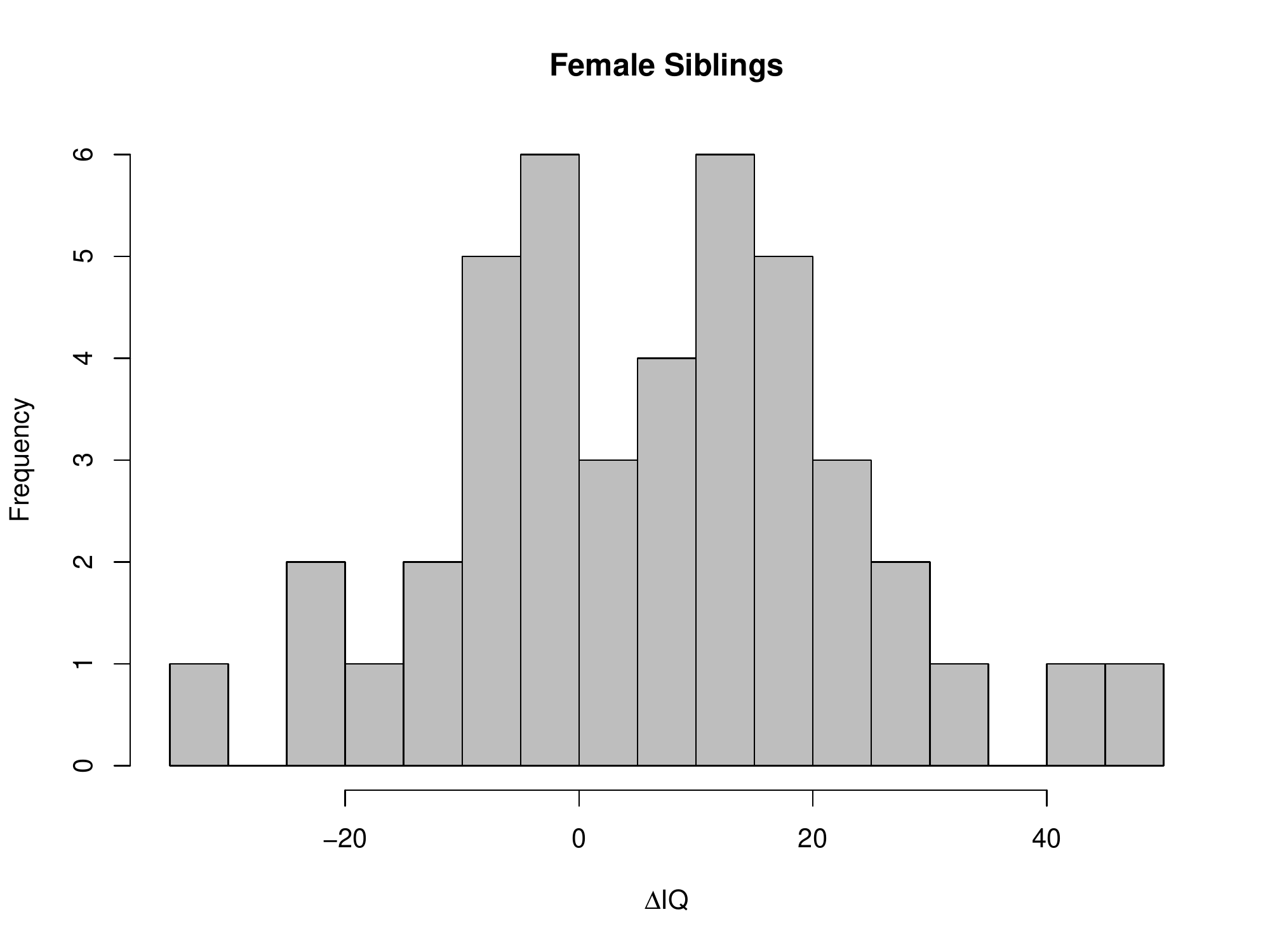}
  \end{tabular}
  \\
  \begin{tabular}{lcccccc}
    \hline
  \textbf{Odds Ratio} & $[1,2)$ & $[2,3)$ & $[3,6)$ & $[6,7)$ & $[7,9)$ & $[9,10)$ \\ 
    \hline
   \textbf{Count} & 165 &   4 &   0 &   1 &   0 &   1 \\ 
     \hline
  \end{tabular}

  \caption{\textit{(left panel)}: Histogram of between-sibling IQ disparities of same-sex male sibling pairs in the WLS study where one sibling attended college and the other did not ($n=128$). \textit{(right panel):} Histogram of between-sibling IQ disparities of same-sex female sibling pairs in the WLS study where one sibling attended college and the other did not ($n=43$). \textit{(bottom panel)}: Table of the estimated increase in pairwise bias due to IQ disparities between siblings measured as an odds ratio.}
\label{fig.histIQpi}
\end{figure}

In many paired studies, sibling or otherwise, hidden biases may strongly influence the results observed for some pairs and more modestly affect others. If the impact of unmeasured confounding were truly heterogeneous in this manner, the conventional sensitivity analysis would be conspicuously conservative. Consider, for example, discrepancies in IQ scores within sibling pairs measured in the WLS where one sibling attended college for at least two years and the other received at most a high school diploma. While existing longitudinal cohort studies rarely contain measures of intelligence \citep{herd2014}, existing evidence suggests that discrepancies in IQ between sibling pairs are strongly predictive of both differences in educational attainment and differences in future income \citep{stanek2011}. In the WLS data, the between-sibling disparity in IQ scores is quite variable across sibling pairs where one sibling attended college and the other did not. The histogram of these college-minus-high school differences is shown in the left panel of Figure \ref{fig.histIQpi} for male sibling pairs and the right panel for female sibling pairs.
Most IQ differences are modest, but a few sibling pairs have large imbalances (e.g. $>40$).

In a sibling study on the returns of schooling where IQ was not recorded, such as Ashenfelter and Rouse's twin study, the maximal bias to which any pair is subject could be materially larger than the typical bias for any sibling pair. Evidence of this pattern's plausibility can be seen in the bottom table of Figure \ref{fig.histIQpi}. The table shows the distribution of the estimated increase in pairwise bias due to IQ disparities between siblings measured as an odds ratio. The numerator of the odds ratio is the predicted maximum odds that the sibling who reported higher income attended college given the reported disparities in IQ while the denominator corresponds to the maximum odds had both siblings had the same IQ.
 (the method for estimating these odds ratios is described in Appendices \ref{sec:B}-\ref{sec:C} ). While the odds ratio in most pairs is close to one, there are a handful of pairs with odds ratios near 2 and two rare cases of odds ratios greater than 6. As far as the `typical' or `expected' pairwise bias is as interpretable a quantity as the worst-case pairwise bias, an \textit{extended} sensitivity analysis of both maximal and expected bias may alleviate concerns that the conventional approach is overly pessimistic while providing a more flexible handling of unobserved bias.

\subsection{Accommodating varying degrees of unmeasured confounding} 

We present an extended sensitivity analysis bounding both the maximal and expected bias for paired studies. The concept of expected bias is made precise in \S \ref{sec.avgcase}. The theoretical foundations and implementation of the extended sensitivity analysis are developed in \S\S \ref{sec:review}- \ref{sec:implement}, while supporting Type I error control and power simulations are presented in \S \ref{sec.sim}. The procedure involves two interpretable parameters, $\Gamma$ and $\bar{\Gamma}\leq \Gamma$, bounding the maximal and expected bias, respectively. At one extreme, setting $\bar{\Gamma} = \Gamma$ recovers the conventional sensitivity analysis for paired studies proposed in \citet[][\S 2]{rosenbaum1987}. At the other, setting $\Gamma=\infty$ for a fixed value of $\bar{\Gamma}$ allows one to bound the average bias while leaving the maximal bias in any given pair unbounded, subsuming the extension presented in \citet[][\S 4]{rosenbaum1987} where the investigator specifies a fraction $\beta$ of the pairs that satisfy a constraint on the maximal bias and allows the remaining pairs to be exposed to potentially unbounded bias.

The procedure builds in two important ways on recent work by \citet{hasegawa2017} that established an exact sensitivity analysis for the sample average bias for paired studies with binary outcomes. First, our procedure accommodates continuous outcomes while providing an asymptotically valid testing procedure for sharp null hypotheses for a large class of test statistics. While generalizing to continuous outcomes corrupts properties unique to McNemar's test statistic utilized in \citet{hasegawa2017}, these difficulties are overcome through a new formulation of the optimization problem necessitated by the sensitivity analysis as a quadratic program. Second, our procedure allows the researcher to bound the expected bias at the level of a superpopulation, rather than the average of the bias at the level of the observed study population, if a superpopulation model is deemed appropriate. This facilitates consonance between superpopulation and finite-sample modes of inference to which the researcher is automatically entitled when only bounding the maximal bias. Actualizing this harmony requires the combination of concentration inequalities with the technique presented in \cite{berger1994} for yielding valid $p$-values by maximizing over a confidence set for nuisance parameters.

To demonstrate the practical consequences of our procedure we return in \S \ref{sec.returns} to the motivating example of returns to schooling. Using the availability of IQ measures in the WLS sibling data, we follow \citet{hsu2013} to estimate the maximal and expected bias under the assumption that inherent cognitive ability is the overwhelming unobserved confounding factor in sibling studies of returns to schooling when IQ measures are not available. We compare standard and extended sensitivity analyses calibrated to these estimates of the sensitivity parameters for Ashenfelter and Rouse's twin study where IQ was not observed.

\section{Sensitivity analysis for paired studies}\label{sec:review}
\subsection{An idealized construction of a paired observational study}\label{sec:ideal}
There are $I$ pairs of individuals. In the $i^{th}$ matched pair one individual receives the treatment, $Z_{ij} = 1$, and the other receives the control, $Z_{ij'} = 0$, such that $Z_{i1} + Z_{i2} = 1$ for each $i$. In practice, the $I$ pairs come into being by minimizing a metric reflective of the within-pair discrepancies between the observed covariates $\x_{ij}$ for the treated and control individuals in a candidate pairing, such that $\bx_{i1}\approx \bx_{i2}$ in the resulting pairs. As an idealization of this practice, we follow \citet{rosenbaum1987} and imagine a generative model where the pairs are constructed, for $i=1,...,I$, by initially drawing, without replacement from an infinite population of treated individuals (that is, conditional upon $Z = 1$), an individual who has an observed covariate $X_i = x_i$. For each $i$, we then sample a control individual from the population of controls with the same value for the observed covariate, i.e. given $Z=0, X= x_i$. Finally, randomly assign indices $(i,1)$ and $(i,2)$ to the two individuals in pair $i$, and let $X_i$ be a random variable denoting the shared value $X_{i1}=X_{i2}$. Despite having a shared value $X_i$, it may be the case that $U_{i1} \neq U_{i2}$ in any pair $i$ for some unobserved covariate $U$. In \S \ref{sec:finitesample}, we describe the extent to which the following methodology applies to finite-sample inference in the absence of a superpopulation.

Under the stable unit-treatment value assumption \citep{rub80}, individual $j$ in matched set $i$ has a potential outcome under treatment, $R_{Tij}$, and under control, $R_{Cij}$ which does not depend on the treatment received by other individuals in the population. The fundamental problem of causal inference is that vector $(R_{Tij}, R_{Cij})$ is not jointly observable. Instead, we observe the response $R_{ij} = R_{Tij}Z_{ij} +R_{Cij}(1-Z_{ij})$, and the observed treated-minus-control paired differences $Y_i = (Z_{i1}-Z_{i2})(R_{i1} - R_{i2})$. Lowercase letters denote realizations of random variables. Let $\mathcal{F}_I = \{(x_{ij}, u_{ij}, r_{Tij}, r_{Cij}),\;\; 1\leq i\leq I, \;\; j = 1,2\}$ be the values of the potential outcomes, measured covariates, and unmeasured covariates for the $2I$ individuals in the observational study at hand. At times it will be convenient to use boldface for vector-valued constants and random variables after the assignment of indices. For example, $\bZ$ represents a vector of length $2I$ with elements $\bZ = (Z_{11}, Z_{12},...,Z_{I2})$, while $\mathbf{R}_{i}$ is a vector of length two with elements $\mathbf{R}_{i} = (R_{i1}, R_{i2})$.

\subsection{Randomization inference under strong ignorability}\label{sec:perm}
The expectation of each paired difference $Y_i$ in the infinite population model of the preceding section is $\E(Y_i \mid X_{ij} = x) = \E(R_{Tij}\mid Z_{ij} = 1, X_{ij} = x) - \E(R_{Cij} \mid Z_{ij}=0, X_{ij} = x)$ which need not equal $\tau(x):= \E(R_{Tij}-R_{Cij}\mid X_{ij} = x)$ without further assumptions on the relationship between the potential outcomes, the observed covariates, and the treatment indicators. A sufficient condition for equality of these expectations, strong ignorability, entails that for any point $x$, 
\begin{align}\label{eq:strongig}(R_T, R_C)\indep Z \mid X,\;\; 0 < \P(Z=1\mid X = x) < 1.
\end{align}

Strong ignorability facilitates far more than equality between $\E(Y_i \mid X_{ij} = x)$ and $\tau(x)$; indeed, it entitles the researcher to use randomization tests akin to those justified in randomized experiments. We consider general hypotheses of the form 
\begin{align*} H_0:\;\;\;& F_{T}(R_{Tij}) = F_{C}(R_{Cij})\;\;\;\forall i,j
\end{align*} for pre-specified functions $F_{T}(\cdot)$ and $F_{C}(\cdot)$. While this form accommodates flexible models for treatment effects, perhaps the most classical specification is the additive treatment effect model where the treatment effect is constant at $\tau$ for all individuals. Under this model $R_{Tij} = R_{Cij} + \tau$, which can be expressed by setting $F_{T}(R_{Tij}) = R_{Tij} - \tau$ and  $F_{C}(R_{Cij}) = R_{Cij}$. From our data alone we observe $F_{ij} = F_{T}(R_{Tij})Z_{ij} + F_{C}(R_{Cij})(1-Z_{ij})$; let $\bF = [F_{11},...,F_{I2}]$.  Under $H_0$, the vectors $\bF_{C} = [F_{C}(R_{C11}),...,F_{C}(R_{CI2})]$ and $\bF_{T} = [F_{T}(R_{T11}),...,$ $F_{T}(R_{TI2})]$ are known to be equal, and hence are entirely specified by the vector of observed responses $\bR$.

Let $t(\bZ, \bF)$ be an arbitrary test statistic that is a function of the treatment indicators $Z_{ij}$ and the observed values $F_{ij}$, and let $\Omega_I = \{\mathbf{z}: z_{i1} + z_{i2} = 1,\;\; 1\leq i\leq I\}$ be the set of $2^I$ possible assignments of individuals to treatment and control in a paired design. Further let $\bof_C$ be the realized value of the random variable $\bF_C$. When $H_0$ holds, $\bof_C$ is fully observed. Under the idealized model in \S \ref{sec:ideal} and under (\ref{eq:strongig}), Theorem 1 of \citet{ros84cp} demonstrates that under the null hypothesis $H_0$, \begin{align} \label{eq:dist}\P\{t(\bZ, \bF) \geq a \mid \cF_I, H_0\} &=\frac{1}{2^I}\sum_{\bz \in \Omega_I} \chi\{t(\bz, \bof_C) \geq a\},\end{align} where $\chi\{A\}$ is an indicator that the event $A$ occurred. Importantly, under $H_0$, the randomization distribution (\ref{eq:dist}) is free of unknown parameters through conditioning on $\cF_I$, and hence can be used directly to facilitate inference on $H_0$.

\subsection{Sensitivity analysis bounding the supremum}\label{sec:sens}
In paired randomized experiments, the physical act of randomization breaks the association between potential outcomes and the intervention and thus justifies both the assumption of strong ignorability and randomization inference through the conditional distribution in (\ref{eq:dist}). Paired observational studies aim to mimic an idealized randomized experiment by creating pairs where individuals are similar on the basis of their observed covariates, $X$, which would similarly facilitate randomization inference through (\ref{eq:dist}) if strong ignorability held. In observational studies, strong ignorability, and in turn belief in (\ref{eq:dist}), turns a statement of fact into a leap of faith due to the potential presence of unobserved factor $U$. That treatment assignment is rarely known to be strongly ignorable given observed covariates $X$ alone necessitates a sensitvity analysis which assesses the robustness of a study's conclusions to factors not included in $X$. A sensitivity analysis operates under the premise that strong ignorability would have been satisfied if an additional pretreatment covariate $U$ had been used in constructing the pairs, that is if for any $x$ and $u$
\begin{align}\label{eq:strongsens}(R_T, R_C)\indep Z \mid (X,U),\;\; 0 < \P(Z=1\mid X = x, U = u) < 1.
\end{align}

A simple model parameterizing the impact of hidden bias presented in \citet[\S 2]{rosenbaum1987} relates $U$ to the assignment mechanism through a parameter $\Gamma = \exp(\gamma) \geq 1$, which constrains the degree to which $U$ can affect the odds of receiving the intervention through a logit model,
\begin{align} \label{eq:logit}
\text{logit}(\P(Z=1\mid X=x, U=u)) &= \kappa(x) + \gamma u ,\;\; 0\leq u \leq 1.
\end{align}
The bounds on $u$ in \eqref{eq:logit} may be viewed as a restriction on the scale of the unobserved covariate that is required for the numerical value of $\gamma$ to have meaning \citep[][Chapter 4]{rosenbaum2002}. Letting $\pi_i = \P(Z_{i1}=1\mid \cF_I)$, (\ref{eq:strongsens}) and (\ref{eq:logit}) then imply $\pi_i = \text{expit}(\gamma(u_{i1} - u_{i2}))$ and $1-\pi_i = \text{expit}(\gamma(u_{i2} - u_{i1}))$. As a result, the model requires that the bound $\pi_i^* = \max\{\pi_i, 1-\pi_i\} = \text{expit}(\gamma|u_{i1}-u_{i2}|) \leq \Gamma/(1+\Gamma)$ holds uniformly for all $i$, but imposes no additional constraints on $\bpi$, and imposes no constraint on the relationship between the unobserved covariate and the potential outcomes. Theorem 1 of \citet{rosenbaum1987}  illustrates that (\ref{eq:strongsens}), (\ref{eq:logit}) and the generative model described in \S \ref{sec:ideal} imply that under a sharp null $H_0$, the distribution $t(\bZ, \bF)$ given $\cF_I$ takes on the modified form
\begin{align}\label{eq:sensdist}\P\{t(\bZ, \bF) \geq a \mid \cF_I, H_0\} = &\sum_{\bz \in \Omega_I}\left[\vphantom{\prod_{i=1}^I}\chi\{t(\bz, \bof_C) \geq a\}\right. \notag \\
  & \left. \times\;\prod_{i=1}^I\text{expit}(\gamma(u_{i1} - u_{i2}))^{z_{i1}}\text{expit}(\gamma(u_{i2} - u_{i1}))^{z_{i2}}\right].\end{align}  At $\Gamma=1\Leftrightarrow \gamma=0$, (\ref{eq:sensdist}) recovers (\ref{eq:dist}), hence representing strong ignorability on the basis of $X$ alone. For $\Gamma>1$, (\ref{eq:sensdist}) depends on the unknown values of $\bu$. A sensitivity analysis proceeds by, for a given value of $\Gamma$, finding bounds on (\ref{eq:sensdist}) by optimizing over the nuisance parameters $\bu \in [0,1]^{2I}$ (or equivalently, optimizing over $\pi_i$ subject to $\pi^*_i \leq \Gamma/(1+\Gamma)$).

We consider test statistics of the form $t(\bZ,\bF) = \bZ^T\bq$ for some function $\bq = \bq(\bF)$, commonly referred to as sum statistics. Examples of sum statistics in paired observational studies include  Wilcoxon's signed rank test and McNemar's test among many others; see \citet[][Chapter 2]{rosenbaum2002} for more on sum statistics. For example, were we to test the null that the treatment effect was constant at zero for all individuals (commonly referred to as Fisher's sharp null hypothesis), then a choice of $q_{ij} = (R_{ij}-R_{ij'})/I = (r_{Cij}-r_{Cij'})/I$ would amount to a choice of the average of the treated-minus-control paired differences in outcomes as the test statistic. In paired studies, arguments parallel to those in \citet[][Chapter 4]{rosenbaum2002} yield that a tight lower bound on (\ref{eq:sensdist}) is found by setting $u_{i1}-u_{i2} = -\text{sign}(q_{i1}-q_{i2})$ for each pair $i$, where $\text{sign}(a)$ is the sign of the scalar $a$. Similarly, a tight upper bound on (\ref{eq:sensdist}) is found by setting $u_{i1}-u_{i2} = \text{sign}(q_{i1}-q_{i2})$ for each $i$. As a further illustration, if one uses the difference in means as the test statistic, the lower (upper) bound is attained through a perfect negative (positive) correlation between the differences in unmeasured covariates and the signs of the treated-minus-control paired differences.

\section{An extended sensitivity analysis}
\label{sec.ext-sens}
\subsection{Average-case unmeasured confounding in paired studies}\label{sec.avgcase}
In \S\S 1.1-1.2, it was argued that large discrepancies in IQ within pairs of siblings, while likely uncommon, would have a large impact on both likelihood of attaining more than a high school degree and on an individual's expected earnings. Were this the only unmeasured confounder, we would then expect most of the values for $\bpi^*$, the maximal probabilities of assignment to treatment within a pair, to not deviate substantially from 0.5, while a few pairs would likely have values for $\pi_i^*$ substantially larger than 0.5. The conventional model for a sensitivity analysis presented in \S \ref{sec:sens} bounds  $\pi_i^*$ by $\Gamma/(1+\Gamma)$ for all pairs. Despite typical discrepancies in IQ likely being small, the smallest value of $\Gamma$ for which (\ref{eq:logit}) and (\ref{eq:sensdist}) hold would be large due to the small number of extremely biased pairs. When utilized in its original form, the sensitivity analysis in \S 2.3 may then paint an overly pessimistic picture of the robustness of the study's findings to unmeasured confounding under this belief, as it cannot account for the `typical' level of unmeasured confounding being different from the worst-case level.

We consider an extension of the conventional sensitivity analysis summarized in \S 2.3 involving two sensitivity parameters, $\Gamma$ and $\bar{\Gamma}$. The first, $\Gamma$, plays a role identical to that of $\Gamma$ in the conventional sensitivity analysis by bounding the supremum of the biased assignment probabilities within a pair. Explicitly, we bound the probabilities of receiving the intervention through a logit form,
\begin{align}\label{eq:logit2}
\text{logit}(\P(Z=1\mid X, U)) &= \kappa(X) + \gamma U,\;\; 0 \leq U \leq 1.
\end{align} That $0 \leq U \leq 1$ trivially implies that for any pair $i$
\begin{align}\label{eq:gmax}
1/2 \leq \text{expit}(\gamma|U_{i1} - U_{i2}|) \leq \frac\Gamma{1+\Gamma}.\end{align} Under (\ref{eq:strongsens}) and the setup of \S \ref{sec:ideal}, (\ref{eq:logit2}) yields that $\Pi_i^* = \max\{\Pi_i,1-\Pi_i\} = \text{expit}(\gamma|U_{i1} - U_{i2}|)  \leq \Gamma/(1+\Gamma)$, where $\Pi_i = \P(Z_{i1} = 1 \mid X_i, \mathbf{U}_i, \mathbf{R}_{Ti}, \mathbf{R}_{Ci}) = \P(Z_{i1}=1 \mid X_i, \mathbf{U}_i)$. We capitalize $U_{ij}$ and $\Pi_i^*$ to emphasize that they themselves are random variables with respect to the superpopulation model in \S 2.1, which would become deterministic by conditioning in $\cF_I$.

The second sensitivity parameter, $\bar{\Gamma}$, serves to bound the \textit{expectation} of the biased probabilities. We define $\mu_{\pi^*} = \E[\Pi_i^*] =\E[\text{expit}(\gamma|U_{i1}-U_{i2}|)]$, and impose that for some value $\bar{\Gamma}$ such that $1\leq \bar{\Gamma}\leq \Gamma$,\begin{align}\label{eq:gbar}1/2  \leq \mu_{\pi^*} \leq \frac{\bar{\Gamma}}{1+{\bar{\Gamma}}}. \end{align} 
Again, this expectation is taken over repeated samples in the idealized setting in \S \ref{sec:ideal}, within which the fixed but unknown values $\pi_i^*$ in our observational study can be modeled as $iid$ realizations of the random variables $\Pi_i^*$. As with the conventional sensitivity analysis, our model makes no assumption about the relationship between the unobserved covariates and the potential outcomes.

Like the conventional sensitivity analysis, our extended procedure solves an optimization problem over a set of nuisance parameters $\bpi$ that satisfy the typical and maximal bias bounds specified in \eqref{eq:gmax} and \eqref{eq:gbar}. Although the population-level bound $\Pi_i^* \le \Gamma/(1+\Gamma)$ implies the corresponding sample level bound $\pi^*_i \le \Gamma/(1+\Gamma)$, the same cannot be said about the corresponding bound on $\mu^*_\pi$. If $\mu^*_\pi \le \bar\Gamma/(1+\bar\Gamma)$, a sample realization $\bar\pi^*$ arbitrarily close to $\Gamma/(1+\Gamma)$ is still possible, however unlikely. To address this, we translate the bound on $\mu^*_\pi$ to a stochastic bound on $\bar\Pi^*$. 

In order to construct this stochastic bound, we consider properties of the random variable $\Pi_i^*$ across draws from the idealized setting in \S \ref{sec:ideal}. From (\ref{eq:gmax}) and $(\ref{eq:gbar})$, we have that for all $i$ $\Pi_i^*$ is bounded above by $\Gamma/(1+\Gamma)$, bounded below by 1/2, and has expectation $\mu_{\pi^*}$ which is itself bounded above by $\bar{\Gamma}/(1+\bar{\Gamma})$. The Bhatia-Davis inequality \citep{bha00} provides the variance upper bound
\begin{align*}\var(\Pi_i^*) &\leq \left(\Gamma/(1+\Gamma)-\mu_{\pi^*}\right)\left(\mu_{\pi^*}-1/2\right)= \nu^2({\Gamma, \mu_{\pi^*}}).\end{align*}
As the $\Pi_i^*$ can further be modeled as $iid$ random variables under the setting being considered, defining $\bar{\Pi}^* = I^{-1}\sum_{i=1}^I\Pi_i^*$, it follows that 
\begin{align*}
\E[\bar{\Pi}^*] = \mu_{\pi^*},\;\; \var(\bar{\Pi}^*) &\leq  \nu^2({\Gamma, \mu_{\pi^*}})/I. \end{align*} If $\var(\Pi_i^*) > 0$ the Central Limit Theorem applies to $\bar{\Pi}^*$, indicating that for any $0 < \beta \leq 0.5$ 
\begin{align}\label{eq:confset} \underset{I\rightarrow \infty}{\lim} \P(\bar{\Pi}^* \in \mathcal{C}_\beta(\Gamma, \mu_{\pi^*}))  \geq 1-\beta, \end{align}
where, because $\bar{\Pi}^* \geq 1/2$ by definition of $\Pi_i^*$
\begin{align}\label{eq:cbeta}\mathcal{C}_\beta(\Gamma, \mu_{\pi^*}) &= \left[1/2,  \mu_{\pi^*}+ I^{-1/2}\Phi^{-1}(1-\beta)\nu({\Gamma, \mu_{\pi^*}})\right],
\end{align}
and $\Phi^{-1}(p)$ is the $p$-quantile of the standard normal distribution. Further, (\ref{eq:confset}) is trivially true if $\var(\Pi_i^*) = 0$, as the upper bound of $\cC_\beta(\Gamma, \mu_{\pi^*})$ is no smaller than $\mu_{\pi^*}$ when $\beta \leq 0.5$. That is, knowledge of $\mu_{\pi^*}$ alone enables the construction of asymptotically valid uncertainty sets for $\bar{\Pi}^*$.
\subsection{Sensitivity analysis bounding the supremum and expectation}\label{subsec:sensext}
Conditional upon $\cF_I$, attention returns to the unmeasured confounders for the individuals in our study population, $\bu$, and the corresponding assignment probabilities $\bpi$. For any value of $\bu$ and value for $\Gamma$, we have that 
\begin{align}\label{eq:sensdist2}\P\{t(\bZ, \bF) \geq a \mid \cF_I, H_0\} = \sum_{\bz \in \Omega_I} \chi\{t(\bz, \bof_C) \geq a\}\prod_{i=1}^I\pi_i^{z_{i1}}(1-\pi_i)^{z_{i2}},\end{align} where $\pi_i = \text{expit}(\gamma(u_{i1}-u_{i2}))$. As the shared notation seeks to emphasize, (\ref{eq:sensdist2}) is precisely the null distribution utilized in (\ref{eq:sensdist}). Here as well as in (\ref{eq:sensdist}), the unmeasured confounders $\bu$, and hence the conditional assignment probabilities $\bpi$, are unknown constants, hindering the desired inference through their presence as nuisance parameters. The approach taken in \S \ref{sec:sens} was to maximize or minimize (\ref{eq:sensdist2}) over  $\bu\in [0,1]^{2I}$ for a given value $\Gamma$, or equivalently over $\pi_i^* \leq \Gamma/(1+\Gamma)$. In what follows, we replace this optimization with one over a subset informed by both $\Gamma$ and $\bar{\Gamma}$ while providing an asymptotically valid level-$\alpha$ test. 

Suppose without loss of generality that we are considering a one-sided, greater than alternative. Let $\mathcal{P}_\beta(\Gamma, \mu_{\pi^*}) = \{\bpi: \bar{\pi}^* \in \mathcal{C}_\beta(\Gamma, \mu_{\pi^*}),\;\; \pi_i^* \leq \Gamma/(1+\Gamma),\;\; 1\leq i\leq I\}$, and consider the following optimization problem:
\begin{align}\label{eq:opt}\underset{\bpi, \mu_{\pi^*}}{\text{maximize}}&\;\;\;\; p(\bpi, \mu_{\pi^*}) = \sum_{\bz \in \Omega_I} \chi\{t(\bz, \bof_C) \geq t(\bZ, \bF)\}\prod_{i=1}^I\pi_i^{z_{i1}}(1-\pi_i)^{z_{i2}}\\
\text{subject to}  
& \;\;\; \bpi \in \mathcal{P}_\beta(\Gamma, \mu_{\pi^*})\nonumber\\
& \;\;\;\;\mu_{\pi^*} \leq \bar{\Gamma}/(1+\bar{\Gamma}).\nonumber
\end{align}
Let $\mathcal{U}_{\beta}(\Gamma, \bar{\Gamma})$ be the set of feasible solutions to (\ref{eq:opt}). Let $\bpi_{\sup, \beta}$ and $\mu_{\sup, \beta}$ be the $\arg\max$ of (\ref{eq:opt}), such that $p(\bpi_{\sup, \beta}, \mu_{\sup, \beta})$ is the tail probability at the solution to (\ref{eq:opt}). If $\bar{\Gamma} < \Gamma$, let $p_{\beta} = p(\bpi_{\sup, \beta}, \mu_{\sup, \beta}) + \beta$; otherwise, let $p_{\beta} = p(\bpi_{\sup, \beta}, \mu_{\sup, \beta})$. 
\begin{proposition}\label{prop:valid}
Suppose we sample $I$ pairs from an infinite population through the procedure in \S 2.1, that treatment assignment is strongly ignorable given $(X,U)$, and that (\ref{eq:gmax}) and (\ref{eq:gbar}) hold at $\Gamma$ and $\bar{\Gamma}\leq \Gamma$ respectively. Then, if $H_0$ is true, for  $0<\beta \leq 0.5$,
\begin{align*}
\underset{I\rightarrow\infty}{\lim}\P(p_{\beta} \leq \alpha\mid H_0) \leq \alpha
\end{align*}
That is, $p_{\beta}$ is an asymptotically valid $p$-value for an extended sensitivity analysis testing $H_0$ with parameters $(\Gamma, \bar{\Gamma})$.

\begin{proof}
We first prove the result for $\bar{\Gamma} < \Gamma$. The proof is similar to that of Lemma 1 in \citet{berger1994}, differing primarily in that the nuisance parameters given $\cF_I$, $\bpi$, are themselves realizations of random variables in the setting of \S \ref{sec:ideal}. Suppose the null hypothesis is true, and let $\mu_0$ be the true value for $\mu_{\pi^*}$. Further, for any set $\cF_I$ let $\bpi_0$ be the true value of $\bpi$. and let $p(\bpi_0, \mu_0)$ be the value of (\ref{eq:sensdist2}) evaluated at $\bpi_0$ and $\mu_0$. 

\begin{align*}
\P(p_{\beta} \leq \alpha) &= \E[\P(p_{\beta} \leq \alpha, \bar{\pi}^*_0 \in \cC_\beta(\Gamma, \mu_0)\mid \cF_I)] +  \E[\P(p_{\beta} \leq \alpha, \bar{\pi}^*_0 \notin \cC_\beta(\Gamma,\mu_0)\mid \cF_I)] \\
&\leq \E[\P(p({\bpi_0}, \mu_0) + \beta \leq \alpha\mid \cF_I)] + \E[\P(\bar{\pi}_0^* \notin \cC_{\beta}(\Gamma, \mu_0)\mid \cF_I)]\\
&= \E[\P(p({\bpi}_0, \mu_0) \leq \alpha -\beta \mid \cF_I)] + \P(\bar{\Pi}^* \notin \cC_{\beta}(\Gamma, \mu_0))\end{align*}  The second line follows from $p(\bpi_0, \mu_0)\leq \sup_{\bpi \in \mathcal{P}_{\beta}(\Gamma, \mu_0)}p(\bpi, \mu_0) \leq p_\beta - \beta$ if $\bar{\pi}^*_0 \in \cC_\beta(\Gamma, \mu_0)$. By validity of (\ref{eq:sensdist2}) at $\bpi_0$ given $\cF_I$, the first term in the third line is less than or equal to $\alpha-\beta$, while (\ref{eq:confset}) illustrates that ${\lim_{I\rightarrow\infty}}\P(\bar{\Pi}^* \notin \cC_{\beta}(\Gamma, \mu_0)) \leq \beta$ for $0 < \beta \leq 0.5$, proving the result for $\bar{\Gamma} < \Gamma$.

If $\bar{\Gamma} = \Gamma$, a solution $\bpi \in \mathcal{U}(\Gamma,\Gamma)$ is $\pi_i = \Gamma/(1+\Gamma)$ if $(q_{i1} > q_{i2})$ and $\pi_i = 1/(1+\Gamma)$ otherwise, which recovers the sensitivity analysis of \S 2.3. Call this solution $\bpi_{\Gamma}$. By arguments in \citet[][Chapter 4]{rosenbaum2002}, this solution yields a tight upper bound for the probability in (\ref{eq:sensdist2}) under the constraint that $\pi^*_i \leq \Gamma/(1+\Gamma)$. Hence, $p(\bpi_{\sup, \beta}, \mu_{\sup, \beta}) =  p(\bpi_{\Gamma}, \Gamma/(1+\Gamma))$ for any $\beta$. At $\bar{\Gamma} = \Gamma$, we simply employ the conventional sensitivity analysis which produces valid $p$-values without an additive increase by $\beta$.
\end{proof}
\end{proposition}

Prior to conducting an extended sensitivity analysis, the practitioner needs to choose a value for $\beta$. A compromise must be made, as $\beta$ acts as a lower bound on the $p$-value reported by the extended sensitivity analysis but larger values of $\beta$ correspond to tighter constraints on $\bar\pi^*$. Accordingly, we recommend that $\beta$ be chosen to be smaller than the precision with which $p$-values are typically reported, but not by much. This recommendation is similar to the guidance given in \cite{berger1994}. 

$p_{\beta}$ yields an asymptotically valid $p$-value for an extended sensitivity analysis with parameters $(\Gamma, \bar{\Gamma})$ because the uncertainty set $\mathcal{C}_\beta(\Gamma, \mu_{\pi^*})$ defined in (\ref{eq:cbeta}) utilizes the Central Limit Theorem. As our random variables $\Pi_i^*$ are bounded, we are entitled to certain distribution-free uncertainty sets based on concentration inequalities which have the desired coverage for all sample sizes $I$; see Appendix \ref{sec:A} for two approaches using Hoeffding's inequality and Bennett's inequality. These sets, used in place of $\mathcal{C}_\beta(\Gamma, \mu_{\pi^*})$ when constructing $\mathcal{P}_{\beta}(\Gamma, \mu_{\pi^*})$, would provide  valid $p$-values for the extended sensitivity analysis through the solution of (\ref{eq:opt}) for all values of $I$. Unfortunately, exact computation of $p_\beta$ through (\ref{eq:opt}) is itself generally intractable, with the additional constraints imposed on the value of $\bar{\pi}$ destroying the properties of the optimization problem solved by the conventional sensitivity analysis which facilitate an exact solution. In \S \ref{sec:implement}, we provide an implementation of our sensitivity analysis valid in large samples by approximating (\ref{eq:sensdist2}) with an appropriate normal distribution, justified under mild conditions. As we employ a normal approximation through our implementation, already implying a large-sample regime, we proceed illustrating the method using the asymptotically valid uncertainty set $\mathcal{C}_\beta(\Gamma, \mu_{\pi^*})$. 

\subsection{On extended sensitivity analyses for observed study populations}\label{sec:finitesample}
Under the superpopulation model described in \S \ref{sec:ideal}, $\Pi_i^*$ is itself a random variable with expectation $\E[\bar{\Pi}^*]$. In randomized experiments and observational studies, the assumption that the individuals in the study arose as a sample from some larger target population is often specious. Such an assumption is not required for inferential statements, as the act of random assignment to intervention itself can form the basis for probabilistic statements and hypothesis tests, endowing randomized experiments with what Fisher referred to as a ``reasoned basis for inference" \citep{fis35}. \citet{ros99} further argues that the most compelling observational studies are not those which are representative of a larger population, but rather those arrived upon through an active choice of the conditions of observation, seeking the ``rare circumstances in which tangible evidence may be obtained to distinguish treatment effects from the most plausible biases" \citep[p. 259]{ros99}.

As (\ref{eq:sensdist}) indicates through conditioning on the study population, $\cF_I$, the classical sensitivity analysis in \S \ref{sec:sens} yields a null distribution for finite-sample inference whose nuisance parameters are the unknown assignment probabilities $\bpi$ for the individuals in the study at hand. The parameter $\Gamma$, which originally served to bound the supremum of the random variables $\Pi_i^*$, also bounds the supremum of the observed values $\pi_i^*$. This yields harmony between inference conducted for the finite study population and inference assuming an infinite population into existence when interest is in the hypothesis $H_0$. Inference given $\cF_I$ is valid on its own, but if a superpopulation model is deemed appropriate, inference given $\cF_I$ yields valid unconditional inference within that framework. 

The motivation for formulating the extended sensitivity analysis with explicit reference to a superpopulation is that while bounds on the supremum of a random variable bound the random variable's realizations, bounds on the expectation of a random variable do not afford bounds in the sample average. The idealized model is used to formulate probabilistic bounds for the sample average $\bar{\Pi}^*$, which then entitle us to a further bound on the average of the realized vector $\bpi^*$. Proposition \ref{prop:valid} indicates that the price to be paid for implementing this bound is the addition of an extra $\beta$ term to the $p$-value, necessitated by the view of $\bpi^*$ as a realization of a random variable. Should a superpopulation model be deemed unreasonable, our model could instead be interpreted as placing a bound on the sample average of the parameters $\bpi^*$, $\bar{\pi}^*$, in the particular observational study being analyzed. This interpretation eliminates the need for both the uncertainty set $\mathcal{C}_\beta(\Gamma, \mu_{\pi^*})$ and the increase in the $p$-value by $\beta$, and an option to consider study population inference is available within our \texttt{R} function. In our particular case study we proceed using superpopulation bounds, as in calibrating the sensitivity parameters in one observational study by means of another one must assume comparability of biases in the two studies.

\subsection{A special case: Binary outcomes}\label{subsec:special}
Although exact computation of $p_{\beta}$ is generally intractable, in one special but common setting it is not. When the outcomes being studied are binary and $t(\bZ,\bF)$ is chosen to be McNemar's test statistic, computing $p_{\beta}$ exactly under Fisher's sharp null $H_0:\; R_{Tij} = R_{Cij}$ becomes a straightforward exercise. Recall that McNemar's test statistic counts the number of pairs where the subject under treatment has a positive outcome and the control subject does not; that is, $t(\bZ,\bF) = \sum_{i=1}^I (Z_{i1}-Z_{i2})(R_{Ci1}-R_{Ci2})/2 + 1/2$ when Fisher's sharp null is true. Since pairs that are not discordant in treatment and outcome do not contribute to McNemar's statistic it is natural to distinguish pairs that are discordant in outcome and those that are not. Let the first $I_{d}$ pairs be the discordant pairs and the last $I_{c}$ be the concordant pairs so that $I = I_{d}+I_{c}$. Furthermore, let the first unit of each discordant pair be the unit with positive outcome, that is $R_{i1}=1$ for $i = 1,\dots,I_{d}$.

For the special case of McNemar's test, let $\mu_m$ be the value of $\mu_{\pi^*}\le \bar\Gamma/(1+\bar\Gamma)$ that maximizes the upper bound of $C_{\beta}(\Gamma,\mu_{\pi^*})$ and let $\bar\pi_{m}$ be the maximized upper bound. Define $\bar\pi_c =1/2$,

\begin{equation*}
\bar{\pi}_{d} = \min\left\{(I\bar\pi_{m}-I_{c}\bar \pi_c)/I_d,\Gamma/(1+\Gamma)\right\},
\end{equation*}  and $\bpi_m = ([\bar{\pi}_{d}\cdot\1_{d},\bar\pi_c\cdot\1_{c}])$, where $\1_{k}$ is a vector of $I_k$ ones.  $(\bpi_m, \mu_m)$ is then a feasible solution to \eqref{eq:opt} that is designed to put as much bias on the discordant pairs as is allowed by the constraints of the optimization problem. Furthermore, since the concordant pairs do not contribute to the test statistic we have that $p(\bpi_m, \mu_m)=\P(B(I_d, \bar{\pi}_d) \ge t(\bZ,\bF)),$ where $B(I_d, \bar{\pi}_d)$ is a Binomial random variable with success probability $\bar{\pi}_{d}$ and $I_{d}$ trials.
Now, let $p_\beta = p(\bpi_m, \mu_m) + \beta$ when $\bar\Gamma < \Gamma$  and let $p_\beta = p(\bpi_\Gamma,\Gamma/(1+\Gamma))$ otherwise.
In the following proposition we show that, in this special setting, an exact solution to \eqref{eq:opt} simply requires computing this Binomial tail probability.

\begin{proposition}\label{prop:special}
Consider a test of $H_0: R_{Tij}=R_{Cij}$ with binary outcomes, and let $t(\bZ,\bF)$ be McNemar's test statistic. Further, let $C_{\beta}(\Gamma,\mu_{\pi^*})$ be an exact, distribution-free $1-\beta$ uncertainty set. Then under the same conditions as Proposition \ref{prop:valid},
\begin{equation*}
  \P(p_\beta \le \alpha\mid  H_0) \le \alpha
\end{equation*}
for any $I$ if $t(\bZ,\bF) \ge I_d \bar\pi_d$. In other words, for any value of $I$, computing a valid $p$-value for an extended sensitivity analysis testing $H_0$ with parameters $(\Gamma,\bar\Gamma)$ reduces to computing the Binomial tail probability $\P(B(I_d, \bar{\pi}_d) \ge t(\bZ,\bF))$.

\begin{proof}
When $\bar\Gamma=\Gamma$, the proof follows immediately from the proof of this case in Proposition \ref{prop:valid}. Hence, we restrict our attention to the case when $\bar\Gamma < \Gamma$. As noted in \S \ref{subsec:sensext}, if we replace $C_{\beta}(\Gamma,\mu_{\pi^*})$ with a distribution-free uncertainty set the optimal solution to \eqref{eq:opt} yields a valid $p$-value for an extended sensitivity analysis for all values of $I$. All that remains to be shown is that $(\bpi_m, \mu_m)$ is the argmax of \eqref{eq:opt}.

Without loss of generality, suppose once again that the first subject of each discordant pair is the unit with a positive outcome, $R_{i1}=1$ for all $i = 1,\dots,I_{d}$.
Let $(\bpi',\mu')$ be a feasible solution of \eqref{eq:opt} and define $\bar{\pi}'_d$ and $\bar{\pi}'_c$ to be the sample average of the maximal assignment probabilities for the discordant and concordant pairs, respectively. $([\bar\pi_d'\cdot\1_d,\bar\pi_c'\cdot\1_c],\mu')$ is clearly also a feasible solution. Then, Theorem 1 in \citet{hasegawa2017} implies that $p([\bar\pi_d'\cdot\1_d, \bar\pi_c'\cdot\1_c],\mu') \ge p(\bpi',\mu')$ when $t(\bZ,\bF) \ge I_d\cdot \bar\pi_d'$. Hence, we need only consider feasible solutions of the form $([\bar\pi_d'\cdot\1_d,\bar\pi_c'\cdot\1_c],\mu')$. An elementary fact about Binomial random variables is that $B(I_d,p_1)$ stochastically dominates $B(I_d,p_2)$ when $p_1 \ge p_2$. By construction, $(\bpi_m,\mu_m)$ yields a feasible solution such that $\bar{\pi}_d \ge \bar\pi_d'$ for all feasible solutions of the form $([\bar\pi_d'\cdot\1_d,\bar\pi'_c\cdot\1_c],\mu')$. Consequently, $p(\bpi_m, \mu_m) \ge p([\bar\pi_d'\cdot\1_d,\bar\pi'_c\cdot\1_c],\mu') \ge p(\bpi',\mu')$ for all feasible solutions $(\bpi',\mu')$ which proves the result for $\bar\Gamma < \Gamma$.

\end{proof}
\end{proposition}

For McNemar's test, the extended sensitivity analysis exhibits an interesting behavior when $\bar{\pi}_d=\Gamma/(1+\Gamma)$: the procedure returns a $p$-value equal to the $p$-value returned by the conventional sensitivity analysis at $\Gamma$ \textit{plus} the extra $\beta$ term. We still pay the cost of specifying a bound on $\E[\Pi_i^*]$ but do not receive the benefit of a tighter constraint on the realization of $\bpi^*$ for discordant pairs. What, exactly, explains this phenomenon? A plausible scenario that may give rise to this behavior is when $I_c >> I_d$, i.e. there are many concordant pairs in the sample of $I$ pairs. In throwing out concordant pairs when using McNemar's statistic, the uncertainty set for $\bar\Pi^*$, the average of $\Pi_i^*$ over all pairs, tells us relatively little about the realized average $\bar\pi^*_d$ over discordant pairs,  reflecting the cost of bounding the marginal expectation $\E[\Pi^*_i]$ instead of the conditional expectation $\E[\Pi^*_i\mid \bR_{Ti}, \bR_{Ci}]$. 

Although this behavior indicates that the extended sensitivity analysis is, in some sense, suboptimal compared to the conventional sensitivity analysis when $I_c >> I_d$, the practical implications are mostly negligible as $\beta$ is chosen to be smaller than the precision with which $p$-values are generally reported. Furthermore, given a choice of $\Gamma$ and conditional on $(I_d,I_c)$, we can a priori determine the value of $\bar\Gamma$ above which the conventional analysis is superior to the extended analysis. Because $(I_d,I_c)$ are known conditional on $\cF_I$, we are not at risk of using the data twice -- once to choose the best test and once to perform that test. Consequently, the resulting sensitivity analyses will still have the appropriate level.

\section{Implementation through quadratic programming}\label{sec:implement}
The test statistics described in  \S \ref{sec:sens} can be represented as the sum of $I$ independent random variables, $\bZ^T\bq = \sum_{i=1}^IT_i$, where $T_i = (q_{i1} + q_{i2})/2 + (Z_{i1}-Z_{i2})(q_{i1}-q_{i2})/2$. This suggests that, under mild regularity conditions, a central limit theorem would be applicable to the distribution of $\bZ^T\bq$ for any value of $\bpi$ in (\ref{eq:sensdist2}) for almost every sample path $\cF_I$. One sufficient condition proposed in the special central limit theorem of \citet[][\S 6.1.2]{haj99} is that, almost surely, 
\begin{align*}
\frac{\sum_{i=1}^I(q_{i1}-q_{i2})^2}{\underset{1\leq i \leq I}{\max}(q_{i1}-q_{i2})^2} \rightarrow \infty,
\end{align*} which requires that no one term $(q_{i1}-q_{i2})^2$ dominates the sum as the number of pairs increases. (An aside: the central limit theorem in \citet[][\S 6.1.2]{haj99} as originally stated applies to sums of the form $\sum_{i=1}^Ia_iX_i$ where $X_i$ are $iid$ random variables; however, the proof can readily be extended to settings where $I \sigma^2 \leq \sum_{i=1}^I \var(X_i) \leq I c \sigma^2$ for $c > 1$ while dropping the requirement of identical distribution, which encompasses the setting of our extended sensitivity analysis). Under a normal approximation, the problem of finding the worst-case $p$-value is equivalent to finding the worst-case deviate. 

Recall that a sensitivity analysis is typically conducted only if the null hypothesis is rejected under the assumption of no unmeasured confounding ($\Gamma=\bar{\Gamma}=1)$, and then proceeds by iteratively increasing the sensitivity parameters until the test fails to reject. Having proceeded to sensitivity analysis only after rejecting the null under no unmeasured confounding, even with one-sided alternatives we can safely consider rejection or failure to reject for sequentially larger values of $\Gamma$ and $\bar{\Gamma}$ based on the minimal squared deviate, an objective function which is preferred for computational reasons alluded to below. Recalling that under (\ref{eq:sensdist2}) we condition on $\cF_I$ and hence treat the vector $\bq$ as fixed, minimizing the squared deviate can be expressed as an optimization problem over the unknown probabilities $\bpi$ as\begin{align}\label{eq:maxsquare}
\min_{\bpi \in \mathcal{U}_\beta({\Gamma, \bar{\Gamma}})} \frac{(t - \E_{\bpi}[\bZ^T\bq\mid \cF_I])^2}{\var_{\bpi}(\bZ^T\bq\mid \cF_I)},\end{align} where $t$ is the observed value of the statistic $t(\bZ, \bF)$, and the expectation and variance are for the test statistic $t(\bZ, \bF)$ under the randomization distribution (\ref{eq:sensdist2}) for a given vector $\bpi$. Under a normal approximation for $t(\bZ, \bF)$, the squared deviate follows a $\chi^2_1$ distribution. By the argument of the previous section, we then reject the null at level $\alpha$ if (\ref{eq:maxsquare}) is greater than or equal to $G^{-1}(1-2(\alpha-\beta))$  for one-sided alternatives or $G^{-1}(1-(\alpha-\beta))$ for two-sided alternatives, where $G^{-1}(p)$ is the $p$ quantile of a $\chi^2_1$ distribution.

The expectation and variance of the contribution of $T_i$ can be expressed as a function of the unknown vector $\bpi$ as
\begin{align}\label{eq:ET} \E_\bpi[T_i \mid \cF_I] &= \bq_i^T\bpi_i\\ \label{eq:VT} \var_\bpi (T_i\mid \cF_I) &= \pi_i(1-\pi_i)(q_{i1}-q_{i2})^2\\
 &= (\bq_i^2)^T\bpi_i - (\bq_i^T\bpi_i)^2\nonumber
\end{align}  where $\bpi_i$ and $\bq_i$ are vectors of length two with elements $\bpi_i = (\pi_{i1},\pi_{i2})$ and $\bq_i = (q_{i1},q_{i2})$, respectively. Suppose without loss of generality that we are considering a one-sided, greater than alternative and that we rejected the null at $(\Gamma, \bar{\Gamma}) = (1,1)$, which implies that $t \geq (2I)^{-1}\sum_{i=1}^I\sum_{j=1}^2q_{ij}$ (i.e. that the observed value of $t$ exceeded its null expectation). Sort each vector $\bq_{i}$ in descending order such that $q_{i1} \geq q_{i2}$. Then, $\var_\bpi(T_i\mid \cF_I) = \var_{\bpi^*}(T_i\mid \cF_I)$ from (\ref{eq:VT}), while from (\ref{eq:ET}) $\E_\bpi[T_i \mid \cF_I] \leq \E_{\bpi^*}[T_i\mid \cF_I] = \bq_i^T\bpi_i^*$ and $(q_{i1}+q_{i2})/2 \leq \E_{\bpi^*}[T_i\mid \cF_I]$. Hence, any feasible solution $\bpi'$ to (\ref{eq:maxsquare}) has an objective value that is no smaller than that of $(\bpi^*)'$, as the variance will be the same while, recalling the iterative nature of a sensitivity analysis, the distance $(t-\E_{(\bpi^*)'}[\bZ^T\bq'\mid \cF_I])^2$ will be smaller than $(t-\E_{\bpi'}[\bZ^T\bq'\mid \cF_I])^2$. Maintaining this ordering of the vectors $\bq_i$, we can express our optimization problem as a function of the maximal probabilities $\bpi_i^*$.

For any candidate $\bpi^*$, we reject under a normal approximation with a one-sided, greater than alternative at level $\alpha-\beta$ if the corresponding squared deviate exceeds its critical value, $G^{-1}(1-2(\alpha-\beta))$ i.e. if $\zeta(\bpi^*, \alpha-\beta) = (t - \E_{\bpi^*}[\bZ^T\bq\mid \cF_I])^2 - G^{-1}(1-2(\alpha-\beta))\var_{\bpi^*}(\bZ^T\bq\mid \cF_I) \geq 0$. We write $\zeta(\bpi^*, \alpha-\beta)$ explicitly as a function of $\bpi^*$ as 
\begin{align*}
\zeta(\bpi^*, \alpha-\beta) &= (t-\bq^T\bpi^*)^2 - G^{-1}(1-2(\alpha-\beta))\sum_{i=1}^I\left((\bq_i^2)^T\bpi^*_i - (\bq_i^T\bpi^*_i)^2\right)
\end{align*}
If we find that $\zeta(\bpi^*, \alpha-\beta) \geq 0$ for all feasible $\bpi^* \in \mathcal{U}_\beta(\Gamma,\bar{\Gamma})$, we can reject the null while asymptotically controlling the size of the extended sensitivity analysis with parameters $(\Gamma, \bar{\Gamma})$ at $\alpha$. The function $\zeta(\bpi^*, \alpha-\beta)$ is convex and quadratic in $\bpi^*$. Meanwhile, we explicitly write the constraints determining membership in $\mathcal{U}_\beta(\Gamma,\bar{\Gamma})$ as
\begin{align}
\label{eq:Sbound} 1/2 \leq \pi_i^* \leq {\Gamma}/({1+\Gamma}),\;\;\; 1\leq i \leq I \\
\label{eq:weird}I^{-1}\sum_{i=1}^I\pi_i^* \leq \mu_{\pi^*} + I^{-1/2}\Phi^{-1}(1-\beta)\left\{\left(\Gamma/(1+\Gamma) - \mu_{\pi^*}\right)\left(\mu_{\pi^*}-1/2\right)\right\}^{1/2}\\
\label{eq:Ebound}\mu_{\pi^*}\leq {\bar{\Gamma}}/({1+\bar{\Gamma}}).
\end{align} For a fixed value of $\mu_{\pi^*}\leq \bar{\Gamma}/(1+\bar{\Gamma})$ the constraints are linear in the unknown maximal probabilites $\pi_i^*$. Hence, for fixed $\mu_{\pi^*}$, the problem $\min_{\bpi^*} \zeta(\pi^*, \alpha-\beta)$ subject to (\ref{eq:Sbound}) and (\ref{eq:weird}) can be written as a quadratic program. With a one-sided alternative, an asymptotically level-$\alpha$ extended sensitivity analysis with parameters $(\bar\Gamma, \Gamma)$ simply requires checking whether the solution to that quadratic program is greater than or equal to zero, rejecting the null if so and failing to reject otherwise. For a two-sided alternative, simply replace $\zeta(\pi^*, \alpha-\beta)$ with $\zeta(\pi^*, (\alpha-\beta)/2)$ to control the level of the procedure at $\alpha$. See \citet{ros92} and \citet{fog16mc} for similar formulations of sensitivity analyses as convex programs.

A minor complication is that for small values of $I$ or for small values for $\beta$, the right-hand side of (\ref{eq:weird}) need not be monotone increasing in $\mu_{\pi^*}$ if $2\bar{\Gamma}/(1+\bar{\Gamma}) \geq \Gamma/(1+\Gamma) + 1/2$, as decreasing $\mu_{\pi^*}$ may lead to an increase in the component dependent on the variance bound which exceeds the corresponding decrease in the additive term $\mu_{\pi^*}$. To remedy this, one can simply find the value for $\mu_{\pi^*}$ over the range $[(\Gamma/(1+\Gamma) + 1/2)/2, \bar{\Gamma}/(1+\bar{\Gamma})]$ which maximizes the right-hand side of (\ref{eq:weird}) through a bisection algorithm, and then proceed with the quadratic program using this single value. If $2\bar{\Gamma}/(1+\bar{\Gamma}) < \Gamma/(1+\Gamma) + 1/2$, the right-hand side of (\ref{eq:weird}) is, subject to (\ref{eq:Ebound}), maximized at $\mu_{\pi^*} = \bar{\Gamma}/(1+\bar{\Gamma})$, so one can proceed by replacing $\mu_{\pi^*}$ with $\bar{\Gamma}/(1+\bar{\Gamma})$ and solving the required quadratic program. Importantly, the method only requires solving a single quadratic program. Quadratic programs can be solved by many free and commercially available solvers; at the author's website \url{http://www.raidenhasegawa.com}, we provide code implementing our method using the \texttt{R} package for the solver \texttt{Gurobi}, which is free for academic use. We also provide options to replace the constraint (\ref{eq:weird}), justified by the Central Limit Theorem, with bounds described in Appendix \ref{sec:A} which are valid for any $I$ through distribution-free concentration inequalities.

\section{Simulations}\label{sec.sim}
\subsection{Type I error control}\label{sec.typeI}
In the following simulations, we demonstrate that the extended sensitivity analysis introduced in \S \ref{sec.ext-sens} has the correct level. We consider two important cases: (1) when no unmeasured bias is present and (2) when the there is unmeasured bias but the sensitivity analysis is conducted at the true values of $\Gamma$ and $\bar{\Gamma}$. In both settings we test Fisher's sharp null that $\tau = 0$ using the difference in means test with desired Type I error control at $\alpha = 0.05$. We set $\beta = \alpha/10 = 0.005$ for conducting the extended sensitivity analysis. The following treatment model, outcome model, and simulation settings were used to conduct the Type I error control simulations:

\begin{enumerate}
  \item{\textbf{Treatment model:}} $\Pi^*_i = 1/2$ with probability $p=2(\Gamma-\bar{\Gamma})/\{(\Gamma-1)(\bar\Gamma+1)\}$ and $\Pi^*_i = \Gamma/(1+\Gamma)$ with probability $1-p$.
  \item{\textbf{Outcome model:}}
  \begin{itemize}
    \item[-]{\textit{unbiased:}} $Y_{i} = \tau\cdot(Z_{i1}-Z_{i2}) + \epsilon_i$ where $\epsilon_i \stackrel{iid}{\sim} \cN(0,1)$,
    \item[-]{\textit{biased:}} $\quad Y_{i} = \tau\cdot(Z_{i1}-Z_{i2}) + \{2\cdot \Chi(\pi_i > 1 - \pi_i) - 1\}\cdot|\epsilon_i|$ where $\epsilon_i \stackrel{iid}{\sim} \cN(0,1)$. 
  \end{itemize}
 
  \item{\textbf{Sensitivity parameters:}}
    \begin{itemize}
      \item[-]{} $\Gamma \in \{1,1.1,1.25,1.5,2\}$,
      \item[-]{} $\bar{\Gamma} \in \{1,1.05,1.1,1.15,1.2,1.25,1.3,1.35,$\\$1.4,1.45,1.5,1.6,1.7,1.8,1.9,2.0\}$,
      \item[-]{} $\bar{\Gamma}\le\Gamma$.
    \end{itemize}
    
  \item{\textbf{Study and simulation size:}} $I = 100$ pairs, $N_{sim} = 5000$ simulations.
\end{enumerate}

In the biased setting, the unit with higher potential outcome under control has higher probability of receiving treatment. When $\Gamma = \bar{\Gamma} = 1$ we use the convention that $p=0/0 = 0$. The value of $p=\P(\Pi^*_i = 1/2)$ was chosen so that the population treatment model satisfies $\E[\bar{\Pi}^*] = \bar\Gamma/(1+\bar\Gamma)$. The results of the simulation study for the biased and unbiased settings are shown in Table \ref{tab.biased-typeI} and the table in Appendix \ref{subsec.unbiased-typeI}, respectively. The extended sensitivity procedure correctly controls the Type I error rate for all pairs of sensitivity parameters $(\Gamma,\bar{\Gamma})$ tested. The first row of each table, where $\bar{\Gamma}=1$, corresponds to tests under the absence of unmeasured confounding. The pairs where $\Gamma = \bar{\Gamma}$ correspond to the conventional worst-case sensitivity analysis. Under the unbiased treatment model, the extended sensitivity analysis is typically more conservative as we increase $\Gamma$ or $\bar{\Gamma}$. In the biased setting, we observe the same pattern as we vary $\Gamma$, but as $\bar{\Gamma}$ approaches $\Gamma$, the level of the extended sensitivity analysis does not decrease monotonically. In fact, at a certain value of $\bar{\Gamma}$, the extended sensitivity analysis becomes less conservative as we approach $\Gamma$. In short, the solution $\boldsymbol{\pi}_{sup,\beta}$ to the optimization problem 
in \eqref{eq:opt} tends to more closely approximate the true allocation $\boldsymbol{\pi}_0$ when $\bar{\Gamma}$ is close to either $1$ or $\Gamma$ in the biased setting.
When $\bar{\Gamma}$ is close to $1$, the feasible set of $\boldsymbol{\pi}$'s is closely bounded around $\boldsymbol{\pi}_0\approx \1\cdot 1/2$. When $\bar{\Gamma}$ is close to $\Gamma$ the true allocation is $\boldsymbol{\pi}_0 \approx \bpi_{\Gamma}$ and the extended sensitivity analysis behaves like the conventional sensitivity analysis, where $\bpi_{\sup,\beta}=\bpi_{\Gamma}$ yields a tight upper bound on the probability in \eqref{eq:sensdist2}. In between these edge cases, when the feasible set of $\bpi$ is relatively large and the trade-off between maximizing expectation and variance is more nuanced, \eqref{eq:opt} may produce solutions $\bpi_{\sup,\beta}$ that yield appreciably more conservative inference than if had we known the true $\bpi_0$.

\begin{table}[ht]
\centering
\begin{tabular}{rrrrrr}
  \hline
  & \multicolumn{5}{c}{$\boldsymbol{\Gamma}$} \\
  $\boldsymbol{\bar{\Gamma}}$ & 1 & 1.1 & 1.25 & 1.5 & 2 \\ 
  \cline{2-6}
  1 & 0.047 & 0.047 & 0.045 & 0.046 & 0.044 \\ 
  1.05 &  & 0.022 & 0.011 & 0.007 & 0.005 \\ 
  1.1 &  & 0.032 & 0.010 & 0.004 & 0.003 \\ 
  1.15 &  &  & 0.012 & 0.002 & 0.002 \\ 
  1.2 &  &  & 0.017 & 0.004 & 0.001 \\ 
  1.25 &  &  & 0.025 & 0.004 & 0.001 \\ 
  1.3 &  &  &  & 0.006 & 0.000 \\ 
  1.35 &  &  &  & 0.009 & 0.001 \\ 
  1.4 &  &  &  & 0.011 & 0.001 \\ 
  1.45 &  &  &  & 0.014 & 0.001 \\ 
  1.5 &  &  &  & 0.025 & 0.001 \\ 
  1.6 &  &  &  &  & 0.003 \\ 
  1.7 &  &  &  &  & 0.004 \\ 
  1.8 &  &  &  &  & 0.006 \\ 
  1.9 &  &  &  &  & 0.011 \\ 
  2 &  &  &  &  & 0.021 \\ 
   \hline
\end{tabular}
\caption{Rejection probability of the true null hypothesis, $H_0:\, \tau = 0$, under the \textit{biased} setting with target Type I error control at $\alpha = 0.05$. The Monte Carlo standard error of these probability estimates is bounded above by $\sqrt{0.05\times 0.95/5000} \approx 0.003$ if the true Type I error rate is 0.05.}
\label{tab.biased-typeI}
\end{table}

\subsection{The power of an extended sensitivity analysis}
The power of a sensitivity analysis quantifies the ability of an observational study design to distinguish treatment effects from unmeasured bias. Formally, it reports for a given study design the probability of rejecting a false null hypothesis for a chosen level $\alpha$ and sensitivity parameter $\Gamma$ under `favorable' conditions, defined in \citet[][Chapter 14]{designofobs}, as the presence of a treatment effect that causes meaningful effects and absence of unmeasured biases. The investigator cannot determine from observable data alone whether or not such favorable conditions hold. An attractive study design would be highly insensitive to unmeasured confounding if she was lucky enough to find herself in this favorable setting.
The power of an extended sensitivity analysis extends this formalism to the triplet $(\alpha,\Gamma,\bar{\Gamma})$. Power simulations for $\alpha = 0.05$ and several pairs of $(\Gamma,\bar{\Gamma})$ are reported in Table \ref{tab.power0.5} and the table in Appendix \ref{subsec.power0.25} for $\tau = 0.5$ and $\tau = 0.25$, respectively. Other than the presence of a `meaningful' treatment effect $\tau$, the simulation settings are identical to the unbiased setting in \S \ref{sec.typeI}. 

Unsurprisingly, the power of the extended sensitivity analysis decreases as $\bar{\Gamma}$ approaches $\Gamma$. If the investigator has reason to believe that unmeasured confounding is heterogeneous and that extreme pairwise unmeasured confounding is possible but relatively rare, the conventional sensitivity analysis is likely unduly conservative. Further, the extended sensitivity analysis allows the investigator to compare the power of competing study designs under different assumptions about the maximal and expected degree of unmeasured confounding.

\begin{table}[ht]
\centering
\begin{tabular}{rrrrrr}
  \hline
  & \multicolumn{5}{c}{$\boldsymbol{\Gamma}$} \\
  $\boldsymbol{\bar{\Gamma}}$ & 1 & 1.1 & 1.25 & 1.5 & 2 \\ 
  \cline{2-6}
  1 & 0.998 & 0.999 & 0.998 & 0.999 & 0.999 \\ 
  1.05 &  & 0.994 & 0.990 & 0.984 & 0.978 \\ 
  1.1 &  & 0.996 & 0.984 & 0.965 & 0.941 \\ 
  1.15 &  &  & 0.977 & 0.947 & 0.896 \\ 
  1.2 &  &  & 0.978 & 0.928 & 0.833 \\ 
  1.25 &  &  & 0.979 & 0.907 & 0.759 \\ 
  1.3 &  &  &  & 0.890 & 0.719 \\ 
  1.35 &  &  &  & 0.884 & 0.664 \\ 
  1.4 &  &  &  & 0.879 & 0.626 \\ 
  1.45 &  &  &  & 0.874 & 0.578 \\ 
  1.5 &  &  &  & 0.882 & 0.541 \\ 
  1.6 &  &  &  &  & 0.505 \\ 
  1.7 &  &  &  &  & 0.478 \\ 
  1.8 &  &  &  &  & 0.463 \\ 
  1.9 &  &  &  &  & 0.472 \\ 
  2 &  &  &  &  & 0.486 \\ 
   \hline
\end{tabular}
\caption{Rejection probability of the false null hypothesis, $H_0:\, \tau = 0$, under the \textit{unbiased} setting with true alternative hypothesis $H_1:\, \tau = 0.5$. The Monte Carlo standard error of these probability estimates is bounded above by $\sqrt{0.5\times 0.5/5000} \approx 0.007$.}
\label{tab.power0.5}
\end{table}

\section{Extended sensitivity analysis for returns to schooling}\label{sec.returns}

\subsection{A model for returns to schooling}
How does going to college affect job earnings? The question and the implications of the many putative answers are important to education policy experts and parents alike. It has been empirically demonstrated that log earnings are nearly a linear function of schooling \citep[see, for instance,][]{card1992}. In the idealized paired observational setting introduced in \S\S \ref{sec:ideal}-\ref{sec:perm} where the treatment condition is attending college for at least two years and the control condition is receiving at most a high school diploma, a hypothesized treatment effect $\tau \times 100$ would describe the percentage increase in earnings associated with attending at least two years of college, the minimum number of years to receive an associates degree. Formally, we consider the multiplicative treatment effect hypothesis $H_{\tau}:\, R_{Tij} = \tau R_{Cij}$ where $(R_{Tij},R_{Cij})$ are potential earnings after attending college or not. Choosing $t(\bZ,\bF)=\bZ^T\bq$ to be the adjusted difference-in-means test comparing log earnings, $q_{ij}$ would take the form $q_{ij} = (\log R_{Tij} - \log R_{Cij'}) - \log(\tau)$ and $q_{ij'} = -q_{ij}$ under $H_\tau$.


Let $X = [X_f,X_s]$ where $X_f$ and $X_s$ are familial and subject level covariates. In an idealized sibling comparison design, the strong ignorability condition in (\ref{eq:strongig}) would hold with respect to $X_f$; that is, if for all $x_f$, 
\begin{equation}
  \label{eq.sibstrong}
  (R_T,R_C)\indep Z \mid X_f,\;\; 0 < \P(Z=1\mid X_f = x_f) < 1.
\end{equation}

If $X_s$ does not affect treatment assignment but does predict potential outcomes, this sibling version of strong ignorability will still hold. For example, in the sibling pairs from the WLS data that we consider in the following section, the age at which income is measured ($AGE$) is different between siblings. If $X_s = AGE$, then it is conceivable that $X_s$ does not affect whether a sibling went to college or not. This would not be the case for people who went to college later in life or whose family characteristics may have changed over time, in which case $AGE$ would be a proxy for those changes. Regardless, model-agnostic adjustment for $X_s$ and $X_f$ can improve the power of the resulting sensitivity analysis \citep{ros02cov}. For example, we can use simple linear regression to adjust for $X$ by replacing $\bq$ with  $(I - H_{X_s})\bq$ where $H_{X_s}$ is the orthogonal projection onto $X_s$ without an intercept.

\subsection{Ashenfelter: Conventional versus extended sensitivity analysis}
To illustrate the differences between the conventional and extended sensitivity analyses, we return to the twin study of \citet{ashenfelter1998} (AR). AR collected survey data on 680 monozygotic twins (340 pairs) attending the Twinsburg Twins Festival in Twinsburg, Ohio during the summers of 1991, 1992, and 1993. We consider the 40 pairs of twins where one twin attend at least two years of college and the other had no more than a high school education, and where both twins were employed at the time of data collection. Assuming no unmeasured confounding, testing Fisher's sharp null $H_0$ yields a $p$-value of $\approx 0.0001$. We obtain a 95\% confidence interval for $\log(\tau)$ of [0.16,0.43] by inverting $H_\tau$ for $\tau \in \bbR_+$ at $\alpha=0.05$ with a two-sided alternative. Exponentiating the endpoints, attending at least two years of college versus receiving at most a high school diploma increased wages by between 17\% and 53\% with 95\% confidence. 

\begin{figure}[ht] 
  \centering
  \includegraphics[scale=0.4]{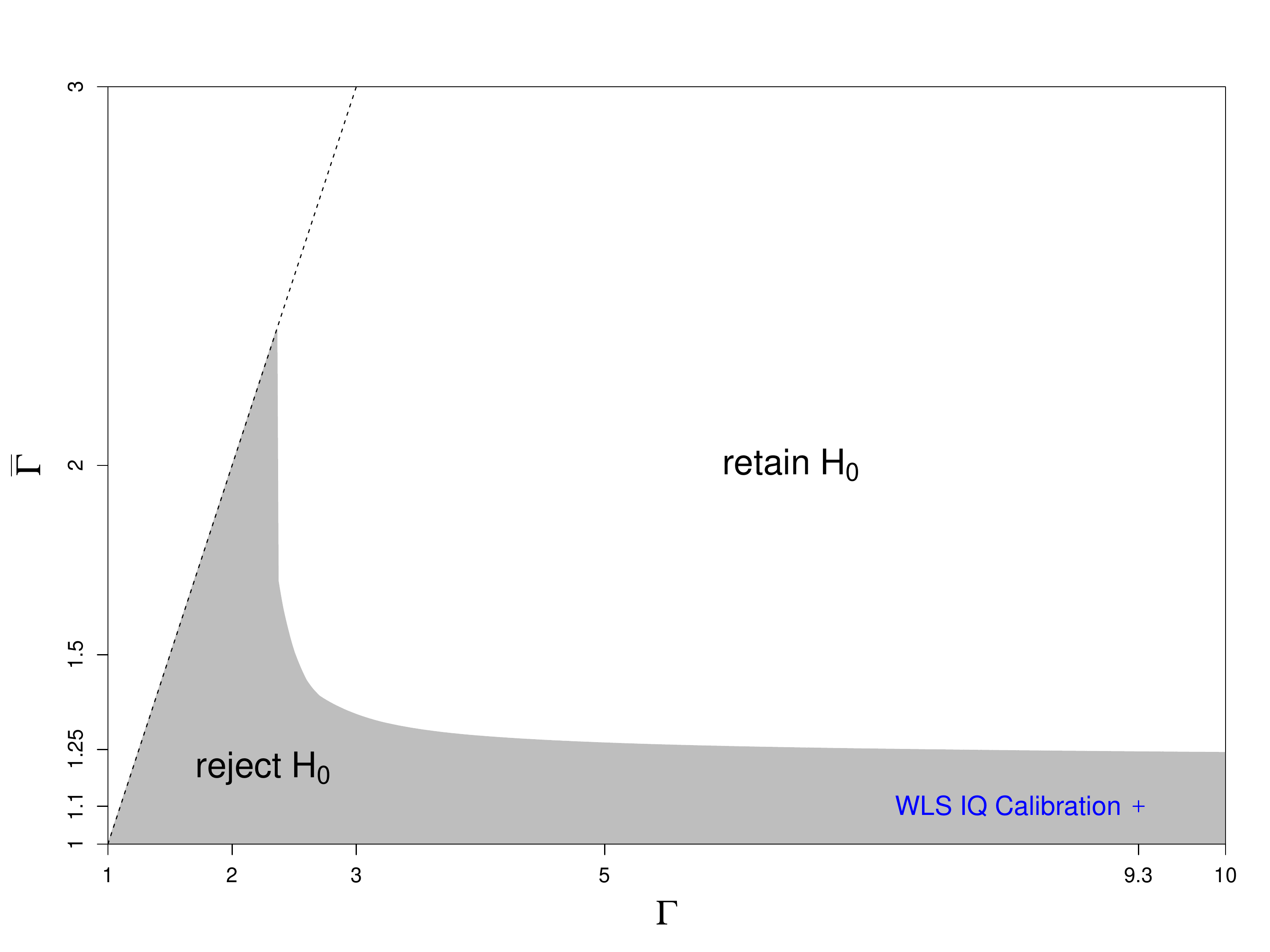}
  \caption{Extended sensitivity curve from the AR study calibrated to the estimates of ability bias from the WLS study (cross). The gray region indicates the sensitivty parameter pairs $(\Gamma,\bar\Gamma)$ for which $H_0$ can still be rejected. The point where the sensitivity curve intersects the $y=x$ line corresponds to the sensitivity value returned by conventional sensitivity analysis ($\Gamma^* \approx 2.36$). The limit of the curve as $\Gamma \to \infty$ corresponds to the sensitivity value returned by the single-parameter sensitivity analysis that bounds the typical bias ($\bar\Gamma^* \approx 1.22$).}  \label{fig.calplot}
\end{figure}

Being a retrospective study neither baseline IQ nor any other intelligence scores were collected, and a critical reader may point to the possible presence of ability bias as a basis to call the conclusions of the study into question. Conducting a sensitivity analysis produces a quantitative rejoinder to this type of criticism in the form of a {\it sensitivity value} $\Gamma^*$ for the conventional analysis and a {\it sensitivity curve} $(\Gamma^*,\bar\Gamma^*)$ for the extended analysis. The sensitivity value is the largest bound on the maximal bias such that the qualitative conclusions of the study do not change (i.e., such that we reject $H_0$). The sensitivity curve is the two-dimensional analog of the sensitivity value and can be seen as the threshold between the gray region (reject $H_0$) and the white region (retain $H_0$) in Figure \ref{fig.calplot}. At the limits of the sensitivity curve, we recover two separate single-parameter sensitivity analyses. The sensitivity value returned by the conventional analysis corresponds to the point where the sensitivity curve intersects the $y=x$ line ($\Gamma^*\approx 2.36$). The limit of the sensitivity curve as $\Gamma \to \infty$ is the sensitivity value of a single-parameter sensitivity analysis that bounds the typical bias ($\bar\Gamma\approx 1.22$).

\subsection{Ability Bias: Cross-study sensitivity analysis calibration}
Without context, the sensitivity curve and values from the Ashenfelter analysis may be difficult to interpret. In response to the critic of the ``equal abilities" hypothesis for twins, we would ideally like to report whether or not the Ashenfelter study is sensitive to plausible patterns of ability bias. One strategy for addressing this is to estimate the bias due to ability from a {\it calibration study} that has a comparable design and information on baseline ability such as IQ. We can then {\it calibrate} the sensitivity analysis to these estimates of $\Gamma$ and $\bar\Gamma$. To implement this {\it cross-study calibration}, we modify the procedure established in \citet{hsu2013} to calibrate sensitivity parameters to observed covariates. In brief, one fits ostensible treatment and outcome models -- for instance, via linear and logistic regression -- and uses the resulting model fits to estimate $\boldsymbol{\pi}^*$, $\bar\Gamma$, and $\Gamma$. The details of this step can be found in Appendix \ref{sec:B}. Calibrating the sensitivity analysis to estimates of ability bias provides the context relevant to the critic's concerns.

To assess the robustness of the AR study to ability bias, we use the sibling data from the WLS study introduced in \S\ref{sec.sibComp} to design a calibration study. We constructed a set of 171 same-sex, full-sibling pairs that received discordant treatment. We let $Z_{ij} = 0$ if sibling $j$ in pair $i$ received 12 or fewer years of education and $Z_{ij} = 1$ if he or she received 14 or more years of education (at least two years of college). Log income for the previous year was collected for WLS participants and their siblings in 1975 and 1977, respectively. To more closely approximate the superpopulation from which the AR twins came, we only consider siblings where both had non-zero income at the time of collection (i.e. were employed). As outlined in the previous section, we let $X_s=AGE$ and use regression to adjust $\bq$ for the age at which income was collected.  This calibration analysis is stylized to some extent to avoid obscuring the primary contribution of our method. Many other subject-level covariates are available for adjustment via regression. A detailed analysis including treatment modification with respect to gender and more thorough covariate adjustment would not preclude the use nor usefulness of our method.

\begin{figure}[ht]
  \centering
  \includegraphics[scale=0.5]{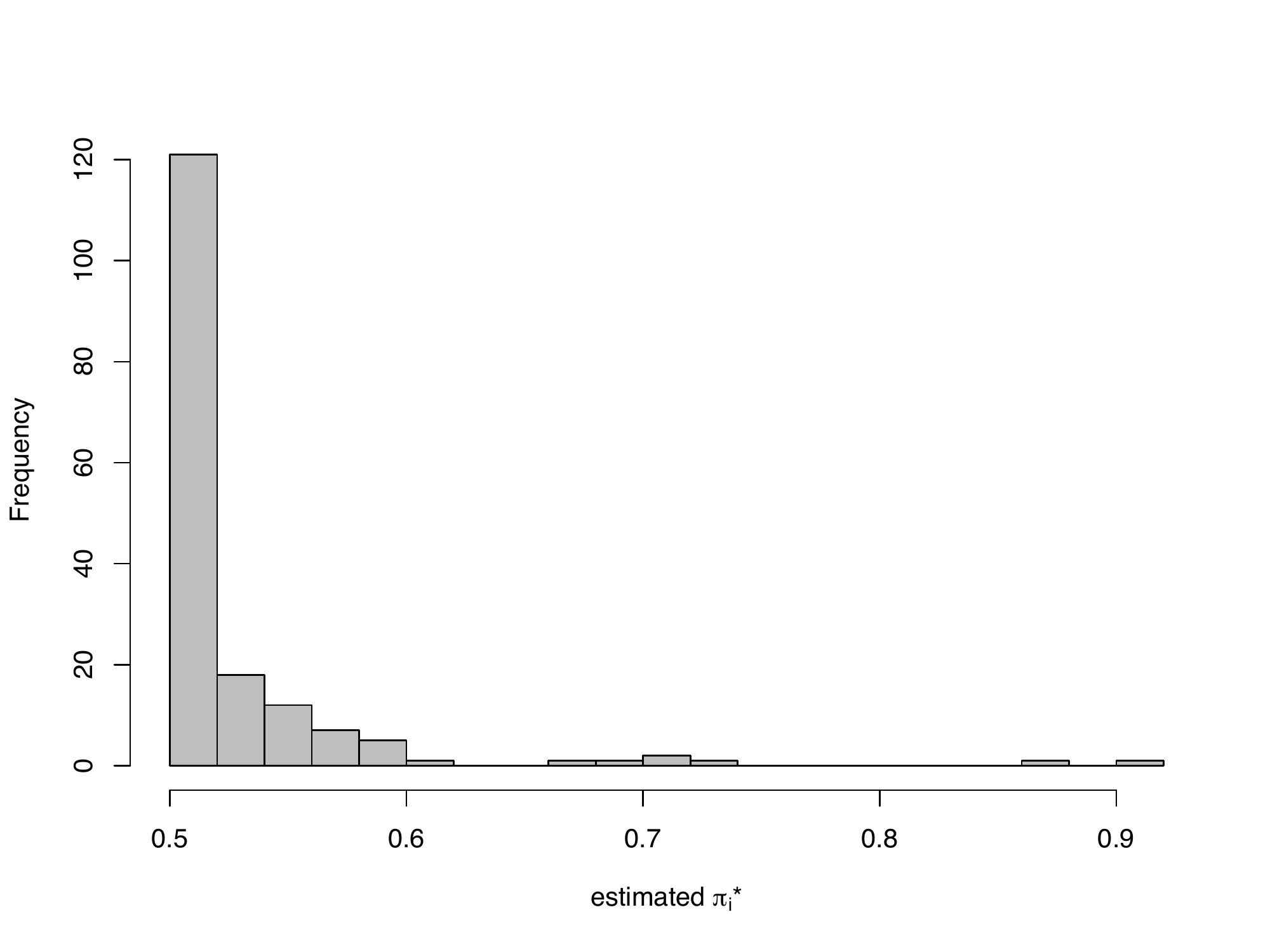}
  \caption{Histogram of $\boldsymbol{\pi}^*$ estimated for 171 same-sex, full-sibling pairs from the WLS study.}  \label{fig.histpi}
\end{figure}

Using the 171 WLS sibling pairs, we estimate that $\Gamma \approx 9.3$ and $\bar\Gamma \approx 1.1$, summarizing the information we have about maximal and typical biases due to IQ disparities. Heterogeneneity of ability bias can explain the considerable difference between these two measures of confounding. The histogram of the estimated $\boldsymbol{\pi}^*$ in Figure \ref{fig.histpi} indicates that most sibling pairs have modest differences in intelligence in high school but in a few rare cases the disparity in sibling IQ exposes pairs to high levels of bias. Calibrating the conventional sensitivity analysis of AR to the WLS study would suggest that our conclusions are likely not robust to plausible patterns of ability bias since $\Gamma^* < 9.3$. However, calibration of the extended sensitivity analysis suggests otherwise. In Figure \ref{fig.calplot}, the WLS IQ calibration point $(9.3,1.1)$ is indicated by the blue cross and falls below the sensitivity curve. The single-parameter sensitivity analysis that bounds the typical bias agrees with the extended analysis that the conclusions are robust to plausible patterns of ability bias ($\bar\Gamma^* \ge 1.1$ ). Incorporating information about the heterogeneity of ability bias by bounding both the maximal and typical biases promotes a less pessimistic assessment of an observational study's robustness to unmeasured confounding. When information on the heterogeneity of potential confounders is available, as in the above cross-study calibration analysis, the extended sensitivity analysis provides a richer picture of the study's robustness to hidden bias.

\subsection{Sensitivity intervals: Interval estimates with hidden bias}
For a fixed bound on the worst-case bias, incorporating heterogeneous bias through the extended sensitivity can also produce narrower \textit{sensitivity intervals} than those attained through the conventional analysis. Representing a natural extension of confidence intervals to inference in the presence of unmeasured confounding,  a $100(1-\alpha)$\% sensitivity interval is constructed by inverting a level-$\alpha$ extended sensitivity analysis with a two-sided alternative at a given pair of values $(\Gamma,\bar{\Gamma})$. Explicitly, let $p_{\beta}(\Gamma,\bar{\Gamma},\tau)$ be the two-sided $p$-value bound returned by the extended sensitivity analysis in \eqref{eq:opt} for particular values of $\Gamma$ and $\bar{\Gamma}$. Then, a $100(1-\alpha)\%$ sensitivity interval can be written as $\cI(\{\tau:\, p_{\beta}(\Gamma,\bar{\Gamma},\tau) \le \alpha\})$, where $\cI(A)$ is the smallest interval containing the set $A$. At $\Gamma=\bar{\Gamma}=1$, the sensitivity interval is simply the corresponding confidence interval found by inverting $H_\tau$ using the randomization $p$-value given in (\ref{eq:dist}) as would be justified in a paired experiment. Setting $\Gamma=\bar{\Gamma} > 1$ returns sensitivity intervals produced through the conventional sensitivity analysis, while setting $\Gamma > \bar{\Gamma} > 1$ employs the extended sensitivity analysis in constructing the sensitivity intervals.

Table \ref{tab.sensint} illustrates the potential for reduced interval lengths through accommodating heterogeneity in unmeasured confounding. It reports 95\% sensitivity intervals for $\log(\tau)$ in the AR study with three pairs of values for $\Gamma$ and $\bar{\Gamma}$. The first, denoted by $\cI_{rand}$, is the 95\% sensitivity interval assuming no unmeasured confounding previously reported in \S 6.2. The second, $\cI_{sup}$, is the 95\% sensitivity interval derived by setting $\Gamma=\bar{\Gamma}=9.3$, the calibrated value of the maximal bias parameter from the WLS study. This is precisely the sensitivity interval that the conventional sensitivity analysis bounding only the worst-case confounding would return. The final interval, $\cI_{ext}$, is the 95\% sensitivity interval setting $\Gamma=9.3$, $\bar{\Gamma}=1.1$ in accord with the calibrated values of the maximal and typical bias from the WLS study. We see that $\cI_{ext}$ is more than 80\% shorter than $\cI_{sup}$. Further, both $\cI_{rand}$ and $\cI_{ext}$ exclude zero while $\cI_{sup}$ does not. The positive finding in the unconfounded setting can be explained away by bias calibrated to the WLS study using the conventional sensitivity model, but not when using the extended sensitivity model. Once again, we see that when it is plausible that the typical bias to which pairs are subject is materially smaller than the worst-case bias, the conventional analysis may be overly pessimistic about how informative the data is.

%
\begin{table}[ht]
  \centering
  \begin{tabular}{rc}
  Interval Type & 95\% Sensitivity Interval\\
    \hline
     $\cI_{rand}$& [0.16,0.43] \\
     $\cI_{sup}$ & [-$0.88$,1.63] \\
    $\cI_{ext}$ & [0.06,0.53] \\
    $100\times(1- |\cI_{ext}|/|\cI_{sup}|)$& 81\% \\
    \hline\\
  \end{tabular}
  \caption{95\% sensitivity intervals for $\log(\tau)$ in the AR study constructed by inverting $H_\tau$ for different values of $\Gamma$ and $\bar{\Gamma}$. $\cI_{rand}$ is the $95\%$ confidence interval for $\log(\tau)$ in the unconfounded setting, $\Gamma=\bar{\Gamma}=1$. $\cI_{sup}$ and $\cI_{ext}$ are 95\% sensitivity intervals derived from the conventional sensitivity analysis and the extended sensitivity analysis respectively. These intervals are formed using the sensitivity parameters calibrated from the WLS data, ($\Gamma, \bar{\Gamma}) = (9.3, 1.1)$. The percentage reduction in interval length from accommodating heterogeneous unmeasured confounding, $100\times(1- |\cI_{ext}|/|\cI_{sup}|)$, is reported in the last row.}
  \label{tab.sensint}
\end{table}

\section{Concluding remarks}\label{sec:conclusion}
While convenient for ease of calculation, the low-dimensional sensitivity analysis bounding the supremum may fail to address specific concerns with unmeasured confounding in certain contexts. \citet{ros09amp} present an amplification of the conventional sensitivity analysis, where the one-dimensional analysis based on $\Gamma$ is mapped to a curve of two-dimensional analyses which simultaneously bound the extent to which differences in unobserved covariates can influence the odds of being treated and the odds of having a higher potential outcome under control by the pair ($\Lambda$,$\Delta$). This amplificiation provides an aid to interpretation, allowing the researcher to posit bounds on the extent to which unmeasured confounding can affect treatment decisions and the outcome variable.  Rather than amplifying the conventional sensitivity analysis, the extended sensitivity analysis provides the researcher a way to further control the distribution of the unmeasured confounders beyond bounding the supremum. In fact, amplification and extension can be viewed as complementary tools available to the researcher. It is straightforward to employ both: the conventional supremum bound $\Gamma$ that appears in the extended sensitivity analysis may be amplified yielding yet an even richer analysis, with $\bar{\Gamma}$ bounding the typical probability that the treated individual in a pair has the larger (smaller) potential outcome under control for greater-than (less-than) alternatives. 

Framing sensitivity analysis in terms of the typical bias is not a new idea, but has been largely unaddressed in the literature; the idea of expected bias appears briefly in \citet{wang2006} in the context of population-level inference for binary outcomes but is not the focus of the paper. In a particular sense, \citet{cor59} anticipated the duality of both amplified and extended sensitivity analyses in their seminal work on sensitivity analysis. In their smoking and lung cancer example, the authors considered a hypothetical hormone $X$ which increases the probability of developing lung cancer among those exposed from $r_2$ to $r_1$ and due to a positive correlation between exposure to $X$ and smoking, appears in a higher proportion among smokers than non-smokers (i.e $p_1 > p_2$). At once, \citet{cor59} captures the spirit of an amplified analysis in specifying how $X$ is related to both treatment assignment and outcome and that of an extended analysis by imagining that hormone $X$ is not completely absent among non-smokers and completely present among smokers, leading to exposure to bias that is heterogeneous across subjects within both groups.

The concept of heterogeneous unmeasured confounding appeared naturally, if not intentionally, in Cornfield's original example. The extended sensitivity analysis introduced in this paper brings this idea into a modern light and provides the researcher with a way to conduct a sensitivity analysis while bounding both maximal and typical biases in matched pair studies. Using two sibling studies on the returns of schooling to income, we demonstrated that a sensitivity analysis bounding the maximal \textit{and} typical bias is both natural and less susceptible to an overly pessimistic view of the study's robustness to hidden bias. When a researcher believes that most, if not all, pairs are exposed to the worst-case bias, our procedure can recover the conventional analysis by setting $\bar\Gamma = \Gamma$. If however, the researcher is worried that some, though few, pairs may be exposed to arbitrarily large biases all is not lost; by letting $\Gamma$ tend to $\infty$ the extended sensitivity analysis recovers a single-parameter sensitivity analysis that bounds the typical bias.

\begin{appendix}
  \setcounter{section}{0}
  \renewcommand{\thesection}{\Alph{section}}
  \section{Construction of Valid Finite-Sample Uncertainty Sets}\label{sec:A}
We now describe the construction of two $100(1-\alpha)\%$ uncertainty sets for $\Pi^*$ valid for any number of pairs $I$. The first is based on Hoeffding's inequality, which implies that the set
\begin{align*}
\mathcal{H}_\beta(\Gamma, \mu_{\pi^*}) = (-\infty, \mu_{\pi^*} + I^{-1/2}\left\{1/2\log(1/\beta)(\Gamma/(1+\Gamma)-1/2)^2\right\}^{1/2}]
\end{align*} satisfies $\P(\bar{\Pi} \in \mathcal{H}_{\beta}(\Gamma, \mu_{\pi^*})) > 1-\beta$ for all values of $I$. 
The second combines Bennett's inequality and the Bhatia-Davis inequality to create the set
\begin{align*}
\mathcal{B}_{\beta}(\Gamma, \mu_{\pi^*}) &= (-\infty, \bar{\mu}_{\pi^*} + b_\beta(\Gamma, \mu_{\pi^*}, I)]\\
b_\beta(\Gamma, \mu_{\pi^*}, I)&= \mathtt{SOLVE}\{a: I^{-1}\log(1/\beta)(\Gamma/(1+\Gamma) - 1/2)^2/\nu^2({\Gamma, \mu_{\pi^*}})= \\& h\left(a(\Gamma/(1+\Gamma) - 1/2)/\nu^2(\Gamma,\mu_{\pi^*})\right)\},
\end{align*}
where $h(x) = (1+x)\log(1+x)-x$. $\mathcal{B}_{\beta}(\Gamma, \mu_{\pi^*})$. This set also satisfies $\P(\bar{\Pi} \in \mathcal{B}_{\beta}(\Gamma, \mu_{\pi^*})) > 1-\beta$ for any $I$ if $\E[\bar{\Pi}^*] = \mu_{\pi^*}$. 

In practice, the upper bound of the set based on Bennett's inequality is smaller than that based on Hoeffding's inequality when $\mu_{\pi^*}$ is far from $(\Gamma/(1+\Gamma) + 1/2)/2$, while the ordering reverses when $\mu_{\pi^*}$ is close to the midpoint. The price paid for this exactness for any $I$ is that the upper bounds for both intervals are larger than those of $\cC_\beta(\Gamma, \mu_{\pi^*})$, the asymptotically valid uncertainty set based on the Central Limit Theorem. 

As noted in the manuscript, the general reliance of our implementation on asymptotic normality reduces the attractiveness of these finite sample uncertainty sets; however, in the case of McNemar's test with binary data, employing either $\mathcal{H}_\beta$ or $\mathcal{B}_\beta$ yields an extended sensitivity analysis for Fisher's sharp null valid for any sample size. \texttt{R} functions to compute these uncertainty set can be found in the file \texttt{multipliers.R} at the author's website \url{http://www.raidenhasegawa.com}.

\section{Constructing the WLS same-sex sibling sample}\label{sec:A2}
Of the 10,317 individuals in the WLS sample, 7,928 had a randomly chosen sibling who was surveyed. Of those 7,928 subjects with sibling data, 2,106 had information about sibling status (i.e. full, half or step siblings) of which 2,004 were full siblings. 1,486 of these sibling pairs were same-sex siblings of which 49.3\% were men. Of the same-sex sibling pairs, in 749 (40.6\% men) both had no more than a high school education, in 265 (64.9\% men) both had at least two years of college education, and in 323 (58.8\% men) one had at most a high school education and the other had at least two years of college education.  The remaining 149 (45.0\% men) same-sex sibling pairs did not meet the definition for our analytic sample because at least one sibling had only one year of college education.  Of the 323 same-sex pairs discordant in educational attainment, 171 (74.9\% men) had complete IQ data and non-zero reported income.

\section{Calibrating Sensitivity Parameters to Disparities in IQ in the WLS Study}\label{sec:B}

We follow a modified version of the calibration strategy introduced in \citet{hsu2013} which involves estimating putative treatment and outcome models as a function of ($X,U$) under $H_0$ via maximum likelihood where the likelihood is marginalized over the unknown confounder $U$. Our modification is as follows: instead of marginalizing over the unobserved covariate we suppose that the only unobserved confounder in the Ashenfelter study is intelligence, which is measured via baseline IQ scores in the WLS study. Consequently, estimating the bias due to IQ disparities using the WLS data permits a cross-study calibration of the Ashenfelter and Rouse sensitivity analysis. 

By definition, $X_f$ is controlled automatically between siblings. We make the stylized assumption that $X_s = AGE$. Further, we assume that $AGE$ does not affect treatment assignment. Finally, we assume that intelligence is the only unmeasured confounder in the Ashenfelter and Rouse study (i.e. $U = IQ$). Under these assumptions, a possible model for treatment assignment is
\begin{equation}
  \P(Z_{ij}=1\mid X_{f,i},\,X_{s,ij},\, U_{ij}) = \frac{\exp(\alpha_{Z,i} + \beta_{Z,IQ}\cdot IQ_{ij})}{1+\exp(\alpha_{Z,i} + \beta_{Z,IQ}\cdot IQ_{ij})}\,.
\end{equation} 

The pair specific intercept $\alpha_{Z,i}$ captures the $X_{f,i}$ effects. We estimate the treatment model using conditional likelihood maximization using the \texttt{R} function \texttt{clogit} in order to avoid bias arising from the fact that the number of $\alpha_i$ to be estimated grows with the sample size. We consider a Gaussian linear model for the outcome
\begin{equation}
  Y_{ij} = \alpha_{Y,i} + \beta_{Y,AGE}\cdot AGE_{ij} + \beta_{Y,IQ}\cdot IQ_{ij} + \epsilon_{ij}\quad \text{such that}\; \epsilon_{ij} \stackrel{iid}{\sim}\cN(0,\sigma^2)\,.
\end{equation}

We estimate the treatment assignment and outcome models using the 171 discordant sibling pairs that we analyze from the WLS study in the paper.

In the Ashenfelter and Rouse twins study, $AGE$ is controlled within twin pairs so we are interested in calibrating the sensitivity parameters to the estimated bias due to $IQ$ disparities alone. Following \citet{hsu2013} we estimate that, controlling for age and assuming that $IQ$ is the only confounding factor, the probability that the sibling that went to college reported a higher income in pair $i$ to be
\begin{align*}
  &\pi_i(\mathbf{IQ}) = \\
  &\frac{\exp\{\hat\beta_{Z,IQ}(IQ_{i1}-IQ_{i2})\}\exp\{(\hat\beta_{Y,IQ}/\hat\sigma^2)(Y_{i(2)}-Y_{i(1)})(IQ_{i1}-IQ_{i2})\} + 1}{[1+\exp\{\hat\beta_{Z,IQ}(IQ_{i1}-IQ_{i2})\}][1+\exp\{(\hat\beta_{Y,IQ}/\hat\sigma^2)(Y_{i(2)}-Y_{i(1)})(IQ_{i1}-IQ_{i2})\}]}
\end{align*}  
where $Y_{i(1)} = \min\{Y_{i1},Y_{i2}\}$ and $Y_{i(2)} = \max\{Y_{i1},Y_{i2}\}$. Define $\bpi(\mathbf{IQ})$ to be the $171\times 1$ vector of $\pi_i(\mathbf{IQ})$. Letting $\bpi^*(\mathbf{IQ}) = \bpi(\mathbf{IQ})$ when $\hat\beta_{Z,IQ}\hat\beta_{Y,IQ}\ge 0$ and $1-\bpi(\mathbf{IQ})$ otherwise, one reasonable set of estimates for ($\Gamma$, $\bar\Gamma$) is
($\pi_{max}/(1+\pi_{max})$, $\bar\pi/(1+\bar\pi)$)
where $\bar\pi = \frac{1}{171}\sum_{i=1}^{171}\pi^*_i(\mathbf{IQ})$ and $\pi_{max} = \sup_i\pi^*_i(\mathbf{IQ})$. It may concern some that $\pi_{max}/(1+\pi_{max})$ is a downwardly-biased estimator of $\Gamma$, but due to sampling variability and possible misspecification of the treatment and outcome models, the calibration is inherently approximate and meant only to act as a guide for the researcher conducting a sensitivity analysis of the Ashenfelter and Rouse study. It should also be noted that since higher $IQ$ does not perfectly predict higher earnings, we find ourselves in a simultaneous sensitivity framework where we simultaneously bound the dependence between $IQ$ and education and between $IQ$ and earnings (see \citet{gas98} for further details). This explains the slightly different definition of $\pi^*_i$ used here than the one found in the paper. Simultaneous sensitivity analysis is closely related to amplified sensitivity analysis, which we discuss briefly in \S \ref{sec:conclusion} of the paper (see \citet{ros09amp} for more details). For our purposes, the simultaneous framework suffices to calibrate $\Gamma$ and $\bar\Gamma$ in the Ashenfelter and Rouse study to the WLS study.

\section{Details of Histogram in Right Panel of Figure \ref{fig.histIQpi}}\label{sec:C}
The figure in the right panel of Figure \ref{fig.histIQpi} in the paper is described as the \textit{[h]istogram of the estimated increase in pairwise bias due to IQ disparities between siblings measured as an odds ratio.} To be specific, and using the notation introduced in Appendix \ref{sec:B}, this is a histogram of
\begin{equation} \frac{\pi_i^*(\mathbf{IQ})}{1-\pi_i^*(\mathbf{IQ})}\left/\frac{\pi_i^*(\0)}{1-\pi_i^*(\0)}\right.
\end{equation} 
for $i = 1,\dots, 171$ where $\pi_i^*(\0) = (1/2)$ is $\pi_i^*$ computed for the sibling pair $i$ had they had same $IQ$ scores.
\newpage
\section{Additional Simulation Results}\label{sec:D}

\subsection{Type I Error Control (\textit{Unbiased})}\label{subsec.unbiased-typeI}

Here we present the results of a simulation of the extended sensitivity analysis under Fisher's sharp null. Table \ref{tab.unbiased-typeI} summarizes the Type I error probabilities at different ($\Gamma,\bar\Gamma$) pairs.

\begin{table}[H]
\centering
\begin{tabular}{rrrrrr}
  \hline
  & \multicolumn{5}{c}{$\boldsymbol{\Gamma}$} \\
  $\boldsymbol{\bar{\Gamma}}$ & 1 & 1.1 & 1.25 & 1.5 & 2 \\ 
  \cline{2-6}
  1 & 0.049 & 0.044 & 0.042 & 0.050 & 0.045 \\ 
  1.05 &  & 0.018 & 0.010 & 0.008 & 0.004 \\ 
  1.1 &  & 0.016 & 0.007 & 0.002 & 0.001 \\ 
  1.15 &  &  & 0.005 & 0.000 & 0.000 \\ 
  1.2 &  &  & 0.003 & 0.000 & 0.000 \\ 
  1.25 &  &  & 0.004 & 0.001 & 0.000 \\ 
  1.3 &  &  &  & 0.000 & 0.000 \\ 
  1.35 &  &  &  & 0.000 & 0.000 \\ 
  1.4 &  &  &  & 0.001 & 0.000 \\ 
  1.45 &  &  &  & 0.000 & 0.000 \\ 
  1.5 &  &  &  & 0.000 & 0.000 \\ 
  1.6 &  &  &  &  & 0.000 \\ 
  1.7 &  &  &  &  & 0.000 \\ 
  1.8 &  &  &  &  & 0.000 \\ 
  1.9 &  &  &  &  & 0.000 \\ 
  2 &  &  &  &  & 0.000 \\ 
   \hline
\end{tabular}
\caption{Rejection probability of the true null hypothesis, $H_0:\, \tau = 0$, under the \textit{unbiased} setting with target Type I error control at $\alpha = 0.05$. The Monte Carlo standard error of these probability estimates is bounded above by $\sqrt{0.05\times 0.95/5000} \approx 0.003$ if the true Type I error rate is 0.05.}
\label{tab.unbiased-typeI}
\end{table}

\subsection{Power at $\tau = 0.25$ (\textit{Unbiased})}

In Table \ref{tab.power0.25} we report the power of the extended sensitivity analysis for $\tau=0.25$ in the unbiased setting. Power is reported over several pairs of ($\Gamma,\bar\Gamma$).

\label{subsec.power0.25}
\begin{table}[H]
\centering
\begin{tabular}{rrrrrr}
  \hline
  & \multicolumn{5}{c}{$\boldsymbol{\Gamma}$} \\
  $\boldsymbol{\bar{\Gamma}}$ & 1 & 1.1 & 1.25 & 1.5 & 2 \\ 
  \cline{2-6}
  1 & 0.694 & 0.677 & 0.677 & 0.694 & 0.683 \\ 
  1.05 &  & 0.544 & 0.462 & 0.391 & 0.338 \\ 
  1.1 &  & 0.528 & 0.363 & 0.282 & 0.188 \\ 
  1.15 &  &  & 0.340 & 0.202 & 0.123 \\ 
  1.2 &  &  & 0.322 & 0.160 & 0.072 \\ 
  1.25 &  &  & 0.333 & 0.132 & 0.046 \\ 
  1.3 &  &  &  & 0.121 & 0.031 \\ 
  1.35 &  &  &  & 0.111 & 0.024 \\ 
  1.4 &  &  &  & 0.110 & 0.019 \\ 
  1.45 &  &  &  & 0.107 & 0.017 \\ 
  1.5 &  &  &  & 0.119 & 0.015 \\ 
  1.6 &  &  &  &  & 0.012 \\ 
  1.7 &  &  &  &  & 0.006 \\ 
  1.8 &  &  &  &  & 0.009 \\ 
  1.9 &  &  &  &  & 0.008 \\ 
  2 &  &  &  &  & 0.010 \\ 
   \hline
   \hline
\end{tabular}
\caption{Rejection probability of the false null hypothesis, $H_0:\, \tau = 0$, under the \textit{unbiased} setting with true alternative hypothesis $H_1:\, \tau = 0.25$ and target Type I error control at $\alpha = 0.05$. The Monte Carlo standard error of these probability estimates is bounded above by $\sqrt{0.5\times 0.5/5000} \approx 0.007$.}
\label{tab.power0.25}
\end{table}

\end{appendix}

\bibliographystyle{imsart-nameyear}
\bibliography{./biblio} 

\label{lastpage}

\end{document}